\documentclass[showpacs,preprintnumbers,amsmath,amssymb]{revtex4}

\usepackage[dvips]{graphicx}
\usepackage{dcolumn}
\usepackage{bm}

\def\mapgeq{\mathbin{\lower.3ex\hbox{$\buildrel>\over{\smash{\scriptstyle\sim}\vphantom{_x}}$}}}
\def\mapleq{\mathbin{\lower.3ex\hbox{$\buildrel<\over{\smash{\scriptstyle\sim}\vphantom{_x}}$}}}
\def\mapgeqeq{\mathbi{\lower.3ex\hbox{$\buildrel>\over{\smash{\scriptstyle\approx}\vphantom{_2}}$}}}
\def\mapleqeq{\mathbin{\lower.3ex\hbox{$\buildrel<\over{\smash{\scriptstyle\approx}\vphantom{_2}}$}}}
\def\Journal#1#2#3#4{{#1} {\bf #2} (#4) #3}
\def\MPL{Mod. Phys. Lett. A}

\def\NPB{Nucl. Phys. B}
\def\NPBOLD{Nucl. Phys.}

\def\PLB{{Phys. Lett.} B}

\def\PLBOLD{Phys. Lett.}
\def\PRL{Phys. Rev. Lett.}
\def\RMP{Rev. Mod. Phys.}
\def\PRD{Phys. Rev. D}

\def\PTP{Prog. Theor. Phys.}
\def\JHEP{JHEP}
\def\EPJ{Euro. Phys. J. C}
\def\JETPUSSR{JETP (USSR)}
\def\ZETP{Zh. Eksp. Teor. Piz.}

\def\IJMP{Int. J. Mod. Phys. A}
\def\IJMPE{Int. J. Mod. Phys. E}
\def\JPG{J. Phys. G}
\def\Erratum{Erratum-ibid}

\begin{document}

\preprint{TOKAI-HEP/TH-0401}

\title{Bilarge Neutrino Mixing and $\mu$ - $\tau$ Permutation Symmetry\\for Two-loop Radiative Mechanism}

\author{Ichiro Aizawa}
 \email{4aspd001@keyaki.cc.u-tokai.ac.jp}

\author{Motoyasu Ishiguro}
 \email{3aspm001@keyaki.cc.u-tokai.ac.jp}

\author{Teruyuki Kitabayashi$^a$}
 \email{teruyuki@post.kek.jp}

\author{Masaki Yasu\`{e}}%
\email{yasue@keyaki.cc.u-tokai.ac.jp}
\affiliation{\vspace{5mm}%
\sl Department of Physics, Tokai University,\\
1117 Kitakaname, Hiratsuka, Kanagawa 259-1291, Japan\\
\\
$^a$
\sl Accelerator Engineering Center \\
Mitsubishi Electric System \& Service Engineering Co.Ltd.\\
2-8-8 Umezono, Tsukuba, Ibaraki 305-0045, Japan}

\date{February, 2004}

\begin{abstract}
The presence of approximate electron number conservation and $\mu$-$\tau$ permutation symmetry of $S_2$ is shown to naturally provide bilarge neutrino mixing.  First, the bimaximal neutrino mixing together with $U_{e3}$=0 is guaranteed to appear owing to $S_2$ and ,then, the bilarge neutrino mixing together with $\vert U_{e3}\vert\ll$1 arises as a result of tiny violation of $S_2$.  The observed mass hierarchy of $\Delta m^2_{\odot}$ $\ll$ $\Delta m^2_{atm}$ is subject to another tiny violation of the electron number conservation.  This scenario is realized in a specific model based on $SU(3)_L\times U(1)_N$ with two-loop radiative mechanism for neutrino masses.  The radiative effects from heavy leptons contained in lepton triplets generate the bimaximal structure and those from charged leptons, which break $S_2$, generate the bilarge structure together with $\vert U_{e3}\vert\ll$1.  To suppress dangerous flavor-changing neutral current interactions due to Higgs exchanges especially for quarks, this $S_2$ symmetry is extended to a discrete $Z_8$ symmetry, which also ensures the absence of one-loop radiative mechanism.
\end{abstract}

\pacs{12.60.-i, 13.15.+g, 14.60.Pq, 14.60.St}
\maketitle

\section{\label{sec:1}Introduction}
Recent observations of terrestrial neutrino oscillations at K2K \cite{K2K} and KamLAND \cite{KamLAND} are proved to be consistent with the observations of atmospheric and solar neutrino oscillations \cite{Kamiokande, SNO}.  These oscillations are characterized by the squared mass differences and mixing angles, which are respectively given by $\Delta m_{atm}^2$ $\sim$ $2\times10^{-3}$ eV$^2$ and $\sin^2 2\theta_{atm} \sim 1$ for atmospheric neutrinos and $\Delta m_\odot^2 \sim 7\times10^{-5}$ eV$^2$ and $\sin^2 2\theta_\odot \sim 0.8$ for solar neutrinos \cite{RecentAnalyses}.  These mixings can be theoretically explained by the oscillations among three known neutrinos, $\nu_{e,\mu,\tau}$, if they are massive \cite{MassiveNeutrino}. In fact, the experimental data indicate that the atmospheric and solar neutrino oscillations can, respectively, originate from the $\nu_\mu$-$\nu_\tau$ mixing and the $\nu_e$-$\nu_\mu$ mixing, which are both almost maximal. The mass hierarchy of $\Delta m_{atm}^2$ $\gg$ $\Delta m_\odot^2$ as well as the large mixing angles suggest that the neutrino mass matrix has almost bimaximal structure \cite{Bimaximal, NearlyBimaximal}. 

To accommodate massive neutrinos with tiny masses of ${\mathcal O}$(10$^{-1}$) eV ($\sim\sqrt{\Delta m_{atm}^2}$ eV), there are two theoretical ideas: one is the seesaw mechanism \cite{Seesaw, type2seesaw} and the other is the radiative mechanism \cite{Zee, Babu}.  While to understand the observed patterns of the neutrino mixings, let us consider, as the first approximation, the maximal neutrino mixings.  To realize these maximal mixings, we often demand specific relations among neutrino masses, for example, $m_{\nu_e\nu_\mu}$=$m_{\nu_e\nu_\tau}$ and/or $m_{\nu_\mu\nu_\mu}$=$m_{\nu_\tau\nu_\tau}$, where $m_{\nu_i\nu_j}$ ($i,j$=$e,\mu,\tau$) stands for the mass for $\nu_i$-$\nu_j$.  To naturally ensure the presence of such specific relations, one may invoke a certain symmetry such as a $U(1)$ symmetry based on $L_e-L_\mu-L_\tau$ ($\equiv L^\prime$) \cite{Lprime} for the maximal solar neutrino mixing or a $\mu$-$\tau$ permutation symmetry for the maximal atmospheric neutrino mixing \cite{ExtendedZeeMuTau,MuTauYasu,GrimusMuTau,MuTau}. 

In this article, we consider the neutrino oscillation based on the radiative mechanism.  It has been argued that the original version of the Zee's one-loop radiative mechanism \cite{Zee} failed to explain the favorable LMA solution for solar neutrinos, namely, the significant deviation from the maximal solar neutrino mixing \cite{ZeeMaximal}.  To get around the difficulty, one has to extend the original Zee's framework \cite{ExtendedZee,ExtendedZeeMuTau} by 1) permitting the second Higgs scalar (introduced in the Zee model) to couple to the leptons, 2) employing a triplet Higgs scalar, 3) including a sterile neutrino or 4) introducing two-loop radiative effects with the absence of one-loop radiative ones.  We are concentrating on examining two-loop radiative effects \cite{Babu} since there are few theoretical models based on two-loop radiative mechanism that explain the observed properties of the current neutrino oscillations.  The advantage of utilizing two-loop radiative effects lies in the fact that Majorana neutrino masses of ${\mathcal O}$(10$^{-1}$) eV can be naturally generated without strong fine-tuning of various couplings.  Denoting the lepton-number violating coupling by $f$ inherent to Majorana neutrino masses, we roughly estimate the one-loop effect to be $fm^2_\tau\mu/16\pi^2M^2$ and the two-loop effect to be $f^2m^2_\tau\mu/(16\pi^2)^2M^2$ for $\mu\sim v_{weak}$ (=$( 2{\sqrt 2}G_F)^{-1/2}$ = 174 GeV) as the mass scale of the standard model and $M\sim$ 1 TeV as the mass scale beyond the standard model.  The estimated mass becomes ${\mathcal O}$(10$^{-1}$) eV for $f\sim 10^{-4}$ in the one-loop effect and $f\sim 10^{-1}$ in the two-loop effect.  The very small magnitude of $f\sim 10^{-4}$ for the one-loop case entails the strong fine-tuning while the magnitude of $f\sim 0.1$ for the two-loop case can be naturally of order of $e$, where $e$ stands for the electric charge. In the present analysis, all the new couplings $f$ beyond the standard-model ones are set to be $\sim e$.  The pattern of the neutrino mixings is also naturally explained by the use of the $\mu$-$\tau$ permutation symmetry.

To respect the power of the permutation symmetry (without fine-tunings of couplings to meet an ``artificial" permutation symmetry), since ordinary charged leptons badly break the permutation symmetry by their masses, new sources, whose interactions can preserve the permutation symmetry, are required.  As the simplest extension of the standard model, we choose an $SU(3)_L\times U(1)_N$ model \cite{331}, which contains heavy leptons in lepton triplets as new sources. $SU(3)_L\times U(1)_N$ models are known to exhibit the intriguing aspect that 
\begin{enumerate}
\item it predicts three families of quarks and leptons if the anomaly free condition on $SU(3)_L\times U(1)_N$ and the asymptotic free condition on $SU(3)_c$ are imposed;
\item the need of the three families of quarks and leptons group-theoretically arises because the anomalies from three families of leptons and one family of quarks that are $SU(3)_L$-triplets are cancelled by those from two families of quarks that are $SU(3)_L$-antitriplets.
\end{enumerate}
Furthermore, these models naturally provide the radiative mechanism for neutrinos of the Majorana type because the standard Higgs doublet together with a Zee scalar specific to the one-loop radiative mechanism is regarded as the Higgs triplet \cite{331RadiativeNu,331Rad,331HeavyERad}.  Our $SU(3)_L\times U(1)_N$ model includes three families of lepton triplets to be denoted by $\psi^i_L$ = ($\nu^i_L$, $\ell^i_L$, $E^i_L$)$^T$ ($i$=$e, \mu, \tau$), where ($\nu^i_L$, $\ell^i_L$)$^T$ is the standard doublet and $E^i_L$ is negatively charged heavy leptons \cite{331HeavyE}.  It can be argued that to activate a two-loop radiative mechanism requires a Higgs scalar of $\xi$=($\xi^{++}$, $\xi^+$, ${\bar \xi}^+$)$^T$ \cite{331HeavyERad}. It will be shown that Eq.(\ref{Eq:OurMnu}) is generated by leading two-loop radiative effects due to the heavy-lepton exchanges while the charged-lepton exchanges add less-dominant two-loop effects that give a slight deviation of $\theta_{atm}$ from the maximal value of $\theta_{atm}=\pi/4$ and a significant contribution on $\theta_\odot$ to give $\sin^22\theta_\odot\sim$0.8. 

In the next section,  we discuss the possible pattern of neutrino mass matrix, which is consistent with the observed neutrino oscillations. In Sec.\ref{sec:3}, we describe the detail of the $SU(3)_L\times U(1)_N$ model including Yukawa and Higgs interactions as well as the symmetry structure summarized in Sec.\ref{sec:2}.  In Sec.\ref{sec:3}, how to generate the neutrino masses and mixings is discussed and the observed properties of neutrino oscillations are shown to be well explained. The final section is devoted to summary.

\section{\label{sec:2}Neutrino Mass Matrix}

To see the appearance of underlying symmetries that approximately describe the observed neutrino oscillations, we show the parameterization of masses and mixing angles in terms of neutrino masses given by a matrix, $M_\nu$,
\begin{equation}
M_\nu =    
    \left(
    \begin{array}{ccc}
         a & b   & c \\
         b & d   & e \\
         c & e   & f\\
    \end{array}
    \right),
	\label{Eq:Mnu}
\end{equation}
on the flavor basis.  The digitalization of $M_\nu$ is performed by the mixing matrix of $U_{MNS}$ \cite{MassiveNeutrino} defined by
\begin{eqnarray}
U_{MNS}&=&\left( {\begin{array}{*{20}c}
   1 & 0 & 0  \\
   0 & {\cos \theta _{23} } & {\sin \theta _{23} }  \\
   0 & { - \sin \theta _{23} } & {\cos \theta _{23} }  \\
\end{array}} \right)
\cdot
\left( {\begin{array}{*{20}c}
   {\cos \theta _{13} } & 0 & {\sin \theta _{13} }  \\
   0 & 1 & 0  \\
   { - \sin \theta _{13} } & 0 & {\cos \theta _{13} }  \\
\end{array}} \right)
\cdot
\left( {\begin{array}{*{20}c}
   {\cos \theta _{12} } & {\sin \theta _{12} } & 0  \\
   { - \sin \theta _{12} } & {\cos \theta _{12} } & 0  \\
   0 & 0 & 1  \\
\end{array}} \right)\\
&=&\left( \begin{array}{ccc}
  \cos\theta_{12}\cos\theta_{13} &  \sin\theta_{12}\cos\theta_{13}&  \sin\theta_{13}\\
  -\cos\theta_{23}\sin\theta_{12}-\sin\theta_{23}\cos\theta_{12}\sin\theta_{13}
                                 &  \cos\theta_{23}\cos\theta_{12}-\sin\theta_{23}\sin\theta_{12}\sin\theta_{13}
                                 &  \sin\theta_{23}\cos\theta_{13}\\
  \sin\theta_{23}\sin\theta_{12}-\cos\theta_{23}\cos\theta_{12}\sin\theta_{13}
                                 &  -\sin\theta_{23}\cos\theta_{12}-\cos\theta_{23}\sin\theta_{12}\sin\theta_{13}
                                 & \cos\theta_{23}\cos\theta_{13}\\
\end{array} \right).
\label{Eq:MNS}
\end{eqnarray}
The masses, $m_{1,2,3}$, and mixing angles, $\theta_{12,23,13}$, for $\theta_\odot$=$\theta_{12}$ and $\theta_{atm}$=$\theta_{23}$, are computed to be
\begin{eqnarray}
&& m_1=c^2_{12}\lambda_1+s^2_{12}\lambda_2-2c_{12}s_{12}A, \quad
 m_2=s^2_{12}\lambda_1+c^2_{12}\lambda_2+2c_{12}s_{12}A,
\nonumber\\
&& m_3=s^2_{23}d+c^2_{23}f+2c_{23}s_{23}e,
\label{Eq:Masses}\\
&& \sin^22\theta_{12} = \frac{8}{8+x^2}, \quad
 \sin^22\theta_{23} = 1-4\Delta^2~\left(c^2_{23}=\frac{1}{2}+\Delta,s^2_{23}=\frac{1}{2}-\Delta\right),
\nonumber \\
&& \tan 2\theta_{13} = 2\frac{s_{23}b+c_{23}c}{m_3-\lambda_1}
\label{Eq:Angles}
\end{eqnarray}
with
\begin{eqnarray}
&& \lambda_1=a, \quad
\lambda_2=c^2_{23}d+s^2_{23}f-2c_{23}s_{23}e, \quad
A = \frac{c_{23}b-s_{23}c}{c_{13}},
\label{Eq:MassesX} \\
&& x = \sqrt{2}\frac{\lambda_2-\lambda_1}{A}, \quad
\Delta = \frac{\sigma\left(f-d\right)+\sqrt{2}\left(b-\sigma c\right)s_{13}}{4e-\sqrt{2}\left(b+\sigma c\right)s_{13}},
\label{Eq:Deviation}
\end{eqnarray}
for $\Delta^2\ll$1, where $c_a=\cos\theta_a$ and $s_a=\sin\theta_a$ for $a$=12, 23, 13, and $\sigma = \pm 1$ for $s_{23}=\pm \vert s_{23} \vert$.

Since $\sin^2\theta_{13}\ll$1 \cite{Ue3} is reported, let us first set $\sin\theta_{13}$=0. For the maximal atmospheric neutrino mixing characterized by $\Delta$=0, the relation of $d$=$f$ (corresponding to $m_{\nu_\mu\nu_\mu}$=$m_{\nu_\tau\nu_\tau}$) is required and can arise as a result of the $\mu$-$\tau$ permutation symmetry while the maximal solar neutrino mixing characterized by $x$=0 arises from $a$=$d$=$e$=$f$=0, indicating the $L^\prime$ conservation.  Our strategy used in this article to realize the bilarge neutrino mixing is to start with the maximal atmospheric neutrino mixing due to the $\mu$-$\tau$ permutation symmetry.  This symmetry allows us to set
\begin{eqnarray}
&& a=0, \quad c= \pm b, \quad f=d,
\label{Eq:ZerothMasses}
\end{eqnarray}
which are subject to
\begin{enumerate}
\item the $L_e$ conservation for $a$=0;
\item a tiny breaking of the $L_e$ conservation for $b\neq 0$;
\item the $\mu$-$\tau$ permutation symmetry for $c$=$\pm b$ and $f$=$d$.
\end{enumerate}
It should be noted that a further possible relation of $d=f=e$ (or $d=f=-e$) may arise from the Majorana mass term of $(\nu_\tau + \nu_\mu)^2$ (or $(\nu_\tau - \nu_\mu)^2$).  

Restricting ourselves to the case of $d=f=e$, thus utilizing $(\nu_\tau + \nu_\mu)^2$, we find the following situations on the masses and mixing angles of neutrinos. In the case of $c_{23}>0$ and $s_{23}>0$, the mass matrix with $c=b$ yields
\begin{eqnarray}
&& m_1=m_2=0, \quad m_3=2d, \quad
 \sin^22\theta_{23}=1, \quad \tan 2\theta_{13}=\sqrt{2} \frac{b}{d},
\label{Eq:ZerothMassesAngles1_positive}
\end{eqnarray}
where $\theta_{12}$ is undetermined, while the mass matrix with $c=-b$ yields
\begin{eqnarray}
&& \vert m_1 \vert = \vert m_2\vert= \sqrt{2}b, \quad m_3=2d, \quad
 \sin^22\theta_{23}=\sin^22\theta_{12}=1, \quad \sin 2\theta_{13}=0.
\label{Eq:ZerothMassesAngles2_positive}
\end{eqnarray}
In the case of $c_{23}>0$ and $s_{23}<0$, the mass matrix with $c=b$ yields
\begin{eqnarray}
&& m_3=0, \quad \sin^22\theta_{23}=1, \quad \sin^22\theta_{12}= \frac{8}{8+x^2}~(x=\frac{2d}{b}), \quad \sin 2\theta_{13}=0,
\label{Eq:ZerothMassesAngles1_negative}
\end{eqnarray}
where $m_1$ and $m_2$ are given by Eq.(\ref{Eq:Masses}) with $\theta_{12}$ given here, while the mass matrix with $c=-b$ yields
\begin{eqnarray}
&&  m_1 = \sin^2\theta_{12}d, \quad m_2= \cos^2\theta_{12}d, \quad m_3=0, \quad
 \sin^22\theta_{23}=1, \quad \sin 2\theta_{13}=1,
\label{Eq:ZerothMassesAngles2_negative}
\end{eqnarray}
where $\theta_{12}$ is undetermined.  The gross feature of the observed neutrino oscillations is compatible with the zero-th order predictions in the cases of $s_{23}>0$ with $b=-c$ \cite{ExtendedZeeMuTau,MuTauYasu} and of $s_{23}<0$ with $c=b$ \cite{GrimusMuTau}.  

We choose the case of $s_{23}>0$ with $c=-b$ that incorporates the bimaximal structure:
\begin{equation}
M_\nu =    
    \left(
    \begin{array}{ccc}
         0 & b   & -b \\
         b & d   & d \\
         -b & d   & d\\
    \end{array}
    \right).
	\label{Eq:OurMnu}
\end{equation}
The hierarchy of $\vert b\vert \ll \vert d\vert$, which is the seed of $\Delta m^2_{\odot}$ $\ll$ $\Delta m^2_{atm}$ realized after $\vert m_1\vert \neq \vert m_2\vert$, originates from the tiny violation of the electron number conservation.  The observed property of $\sin^22\theta_\odot \sim 0.8$ is explained by tiny violation of $S_2$. In the next section, we realize this starting neutrino mass texture in a $SU(3)_L\times U(1)_N$ gauge model, where mass texture of heavy leptons as the third member of a lepton triplet plays an essential r\^{o}le.

\section{\label{sec:3}Model}
The particles in our model are specified by quantum numbers placed as $(SU(3)_L, U(1)_N)$. The hypercharge $Y$ and the electric charge $Q$ are given by $Y$=$\lambda^8/\sqrt{3}+N$ and $Q=(\lambda^3+Y)/2$ respectively, where $N/2$ is the $U(1)_N$ number and $\lambda^a$ are the Gell-Mann matrices with Tr$(\lambda^a \lambda^b)=2\delta^{ab} (a,b=1,2,...,8)$. Leptons are assigned to be:
\begin{eqnarray}
\psi^i_L = \left( \nu^i,\ell^i,E^i\right)_L^T : \left( \textbf{3}, -2/3 \right), \quad
\ell^{e,\mu,\tau}_R : \left( \textbf{1}, -1 \right),  \quad  
E^{e,-,+}_R    : \left( \textbf{1}, -1 \right),
\label{Eq:leptons}
\end{eqnarray}
where $E_R^j$ for $j$ = ($e,-,+$) are the mass eigenstates of the negatively charged heavy leptons and the superscripts, $\pm$, of $E_{\pm R}$ represent the chiral partners of the $\tau \pm \mu$ states of $E_{\pm L}$ to be defined by their Yukawa interactions while quarks are assigned to be:
\begin{eqnarray}
&&Q^{e}_L=\left(d,-u,u^{\prime} \right)_L^T
       :\left(\textbf{3}^\ast,1/3\right),
Q^{\mu}_L=\left(s,-c,c^{\prime}\right)_L^T
       :\left(\textbf{3}^\ast,1/3\right),
Q^{\tau}_L=\left(t,b,b^{\prime}\right)_L^T
	   :\left(\textbf{3},0 \right),~
\nonumber \\
&&u_R,c_R,t_R      : \left( \textbf{1}, 2/3 \right),~ 
d_R,s_R,b_R      : \left( \textbf{1},-1/3 \right),~ 
u^{\prime}_R,c^{\prime}_R : \left( \textbf{1}, 2/3 \right),~
b^{\prime}_R : \left( \textbf{1},-1/3 \right).
\label{Eq:Quarks}
\end{eqnarray}
Another notations for $Q^i$, $Q^i$=$\left(d^i,u^i,u^{\prime i} \right)_L^T$ ($i$=$e,\mu$) and $Q^\tau$=$\left(u^\tau,d^\tau,d^{\prime \tau} \right)_L^T$, and for the corresponding right-handed quarks are also used.  All gauge anomalies are cancelled by these quarks and leptons \cite{331HeavyE}. Higgs scalars are assigned to be:
\begin{eqnarray}
\eta = \left(\eta^0,\eta^-,{\bar \eta}^-  \right)^T : \left( \textbf{3}, -2/3 \right), \quad
\rho = \left(\rho^+,\rho^0,{\bar \rho}^{0} \right)^T : \left( \textbf{3}, 1/3 \right), \quad
\chi = \left(\chi^+,{\bar \chi}^0,\chi^{0} \right)^T : \left( \textbf{3}, 1/3 \right),
\label{Eq:higgs}
\end{eqnarray}
which develop the following vacuum expectation values (VEV's):
\begin{eqnarray}
\langle 0 \vert\eta\vert 0 \rangle = \left(v_\eta,0,      0      \right)^T, \quad
\langle 0 \vert\rho\vert 0 \rangle = \left(0,     v_\rho, 0      \right)^T, \quad
\langle 0 \vert\chi\vert 0 \rangle = \left(0,     0,      v_\chi \right)^T,
\label{Eq:VEVforEtaRhoChi}
\end{eqnarray}
where the orthogonal choice of these VEV's will be guaranteed by appropriate Higgs interactions introduced in Eq.(\ref{Eq:higgsV}).

Since the model respects the $\mu$-$\tau$ permutation symmetry of $S_2$, in order to generate the phenomenologically consistent charged-lepton masses that do not respect the $\mu$-$\tau$ permutation symmetry, the model employs the following $S_2$-antisymmetric Higgs scalars:
\begin{eqnarray}
\rho^\prime = \left(\rho^{\prime +},\rho^{\prime 0},{\bar \rho}^{\prime 0} \right)^T : \left( \textbf{3}, 1/3 \right), \quad
\rho^{\prime\prime} = \left(\rho^{\prime\prime +},\rho^{\prime\prime 0},{\bar \rho}^{\prime\prime 0} \right)^T : \left( \textbf{3}, 1/3 \right),
\label{Eq:ExtraHiggs}
\end{eqnarray}
as well as $S_2$-symmetric Higgs scalar, $\chi^\prime$:
\begin{eqnarray}
\chi^\prime = \left(\chi^{\prime +},{\bar \chi}^{\prime 0},\chi^{\prime 0} \right)^T : \left( \textbf{3}, 1/3 \right)
\label{Eq:ExtraHiggsChi}
\end{eqnarray}
with
\begin{eqnarray}
\langle 0 \vert\rho^\prime\vert 0 \rangle = \left( 0, v_{\rho^\prime},    0      \right)^T, \quad
\langle 0 \vert\rho^{\prime\prime}\vert 0 \rangle = \left( 0, v_{\rho^{\prime\prime}},    0      \right)^T, \quad
\langle 0 \vert\chi^\prime\vert 0 \rangle = \left(0,     0,      v_{\chi^\prime} \right)^T.
\label{Eq:VEVforEtaRhoChiExtra}
\end{eqnarray}
The two-loop radiative mechanism can be initiated by introducing three Higgs scalars denoted by $\xi^\prime$, $k^{++}$ and $k^{\prime ++}$ as $S_2$-symmetric states and $\xi$ as an $S_2$-antisymmetric state:
\begin{eqnarray}
\xi = \left(\xi^{++},\xi^{+},{\bar \xi}^{+} \right)^T : \left( \textbf{3}, 4/3 \right), \quad
\xi^\prime = \left(\xi^{\prime ++},\xi^{\prime +},{\bar \xi}^{\prime +} \right)^T : \left( \textbf{3}, 4/3 \right),
k^{++},k^{\prime ++} : \left( \textbf{1}, 2 \right),
\label{Eq:ExtraHiggsXi}
\end{eqnarray}
where $\xi^\prime$ has $L_e$=1 and others have $L_e$=0. 

Since our model contains quarks with the same charge, whose mass terms can be generated by $\rho$ and $\chi$ between $Q_L^e$ and down-type quarks and by $\rho^\dagger$ and $\chi^\dagger$ between $Q_L^{\mu,\tau}$ and up-type quarks, dangerous flavor-changing-neutral-currents (FCNC) interactions are generally induced at the phenomenologically unacceptable level \cite{FCNCSU3}.  To avoid these interactions, Yukawa interactions must be constrained such that a triplet quark flavor gains a mass from only one Higgs field \cite{FCNC}.  The lepton sector also contains the similar FCNC problem because $\ell^i$ ($i$=$e,\mu,\tau$) and $E^i$ ($i$=$e,\pm$) has the same charge.  It is known that to impose such a constraint on FCNC is readily achieved by introducing a certain discrete symmetry.  We combine the $S_2$-permutation symmetry and the discrete symmetry for the FCNC-suppression into a $Z_8$ symmetry.  Listed in Table \ref{Tab:particlesAndSymmetries} are the quantum numbers of $S_2$, $Z_8$, $L$ and $L_e$ for leptons and Higgs scalars in our discussions, where $\psi_{\pm L}=(\psi_L^\tau \pm \psi_L^\mu)/\sqrt{2}$, $\ell_{\pm R}=(\tau_R \pm \mu_R)/\sqrt{2}$ and $S_2$ is shown for comparison.  In this table, we have omitted the quark sector of the model since $Z_8$ of quarks can be easily adjusted to respect the $Z_8$ conservation for the given $Z_8$ of Higgs scalars to reach quark interactions in Eq.(\ref{Eq:Yukawa}).

The Yukawa interactions for leptons are now described by $\mathcal{L}_Y$:
\begin{eqnarray}
-\mathcal{L}_Y &=& 
	\overline{\left( \psi_{+ L} \right)^c}\left(
	f_{\xi^\prime}\psi_L^e \xi^\prime
	+
	f_{\xi} \psi_{-L} \xi
	\right)
\nonumber \\
&&	+
	\left[
		f^{+\ell}_k\overline{\left( \ell_{+R} \right)^c} \ell_{+R}
		+
		\frac{1}{2}f^{-E}_k\overline{\left( E_{-R} \right)^c} E_{-R}
	\right]k^{++}
	+
	\left[
		f^{-\ell}_{k^\prime}\overline{\left( \ell_{-R} \right)^c} \ell_{-R}
		+
		\frac{1}{2}f^{+E}_{k^\prime}\overline{\left( E_{+R} \right)^c} E_{+R}
	\right]k^{\prime ++}
\nonumber \\
&& 	+f_\ell  \overline{\psi_L^e}\rho e_R 
   	+ \overline{\psi_L^+}
	\left(
		f^+_\ell \rho \ell_{+R} + f^-_\ell \rho^\prime \ell_{-R}
	\right)
   	+ \overline{\psi_L^-}
	\left(
		g^+_\ell \rho^{\prime\prime} \ell_{+R} + g^-_\ell \rho \ell_{-R}
	\right)
\nonumber \\
&& 	
	+ f_E\overline{\psi_L^e} \chi E^e_R
	+ f^+_E \overline{\psi_{+L}}\chi E_{+R}
   	+ f^-_E \overline{\psi_{-L}}\chi^\prime E_{-R}
\nonumber \\
&& 	
	+ \sum_{i=e,\mu}
	\overline{Q^i_L}
	\left( 
		\eta^c D^{\prime i}_R
		+ \rho^c U^i_R 
		+ \chi^c U^{\prime e}_R
	\right)
	+ \overline{Q^\tau_L} 
	\left( 
		\eta U^\tau_R
		+\rho D^\tau_R
		+ \chi D^{\prime \tau}_R
	\right)
   	+ (h.c.)
\label{Eq:Yukawa}
\end{eqnarray}
with
\begin{eqnarray}
&& 
	U^i_R= \sum_{j=e,\mu,\tau}f^i_{uj}u^j_R, \quad
	D^i_R= \sum_{j=e,\mu,\tau}f^i_{dj}d^j_R, \quad
	U^{\prime i}_R= \sum_{j=e,\mu}f^i_{u^\prime j}u^{\prime j}_R, \quad
	D^{\prime i}_R = f^i_{d^\prime \tau}d^{\prime\tau}_R, 
\label{Eq:QuarkFields}
\end{eqnarray}
where $f$'s and $g$'s denote the Yukawa couplings. The Higgs interactions are given by Hermitian terms composed of $\phi_\alpha^\dagger \phi_\beta$ $(\phi=\eta, \rho, \rho^\prime, \rho^{\prime\prime}, \chi, \chi^\prime, \xi, \xi^\prime, k^{++})$, which include the potential terms of $V_{\eta\rho\chi,\rho\chi}$ 
\begin{eqnarray}
&V_{\eta\rho\chi} = \lambda_{\eta\rho} \vert \eta \times \rho \vert^2
      + \lambda_{\rho\chi} \vert\rho \times \chi\vert^2
      + \lambda_{\chi\eta} \vert\chi \times \eta\vert^2,
\label{Eq:Orthgonal}\\
& V_{\rho\chi} = 
	\lambda_{\rho\rho^\prime} \vert \rho^\dagger \rho^\prime \vert^2 
	+ 
	\lambda_{\rho\rho^{\prime\prime}} \vert \rho^\dagger \rho^{\prime\prime} \vert^2
	+
	\lambda_{\chi\chi^\prime}\vert \chi^\dagger \chi^\prime \vert^2
\label{Eq:OrthgonalExtra}
\end{eqnarray}
with the definition of $(a \times b)^\alpha$ $\equiv$ $\epsilon^{\alpha\beta\gamma}a_\beta b_\gamma$ and by non-Hermitian terms in 
\begin{eqnarray}
V  &=&
    \frac{1}{2}\lambda_{\eta\chi\xi\chi^\prime}
	\left[
		(\eta^\dagger\chi)(\xi^\dagger\chi^\prime)
		+
		(\eta^\dagger\chi^\prime)(\xi^\dagger\chi)
	\right]
	+
    \lambda_{\eta\rho \xi\rho} (\eta^\dagger\rho)(\xi^\dagger\rho)
	+
    \frac{1}{2}\lambda_{\eta\rho^\prime\xi\rho^{\prime\prime}}
	\left[
		(\eta^\dagger\rho^\prime)(\xi^\dagger\rho^{\prime\prime})
		+
		(\eta^\dagger\rho^{\prime\prime})(\xi^\dagger\rho^\prime)
	\right]
\nonumber\\
	&&+
	\mu_{\xi\eta k}\xi^\dagger\eta  k^{++}+ \mu_{\xi^\prime \eta k^\prime}\xi^{\prime\dagger}\eta k^{\prime ++}
    + (h.c.),
\label{Eq:higgsV}
\end{eqnarray}
where $\mu$'s and $\lambda$'s denote mass scales and coupling constants, respectively. These Yukawa and Higgs interactions are invariant under 
\begin{itemize}
\item the $Z_8$-transformation, which is spontaneously broken;
\item the $L_e$-transformation, which is not spontaneously broken but explicitly broken by $\xi^{\prime\dagger}\eta k^{\prime ++}$.
\end{itemize}
The $S_2$-permutation symmetry is preserved in the Yukawa interactions but explicitly broken by the Higgs interactions. The $b$- and $c$-terms in Eq.(\ref{Eq:Mnu}) can be generated by interactions containing $\xi^{\prime\dagger}\eta k^{\prime ++}$.  All other interactions are forbidden.  We note that 
\begin{itemize}
\item the orthogonal choice of VEV's of $\eta$, $\rho$ and $\chi$ as in Eq.(\ref{Eq:VEVforEtaRhoChi}) is supported by $V_{\eta\rho\chi}$ if all $\lambda$'s are negative because $V_{\eta\rho\chi}$ gets lowered if $\eta$, $\rho$ and $\chi$ develop VEV's and one can choose VEV's such that $\langle 0\vert\eta_1 \vert 0\rangle$ $\neq$ 0, $\langle 0\vert\rho_2 \vert 0\rangle$ $\neq$ 0 and $\langle 0\vert\chi_3 \vert 0\rangle$ $\neq$ 0;
\item the choice of VEV's of $\rho^\prime$, $\rho^{\prime\prime}$ and $\chi^\prime$ as in Eq.(\ref{Eq:VEVforEtaRhoChiExtra}) is supported by $V_{\rho\chi}$ if all $\lambda$'s are negative; 
\item  a $(\eta^\dagger\chi)(\xi^\dagger\rho)$-term, which is absent in $V$, would cause one-loop radiative effects such as in Fig.\ref{Fig:WouldBeOneLoop} and similarly for the terms with $\rho\rightarrow\rho^\prime,\rho^{\prime\prime}$ and $\chi\rightarrow\chi^\prime$;
\item the suppression of FCNC for quarks is ensured by the absence of $\chi D^e_R$ and $\rho D^{\prime e}_R$ for $Q^e_L$, of $\chi^\ast U^i_R$ and $\rho^\ast U^{\prime i}_R$ for $Q^i_L$ and of $\chi \ell^i_R$ and $\rho E^i_R$ for $\psi^i_L$.
\end{itemize}
Our model also respects
\begin{itemize}
\item the $L$-conservation if $\lambda_{\eta\chi\xi\chi^\prime}$, $\lambda_{\eta\rho \xi\rho}$ and $\lambda_{\eta\rho^\prime\xi\rho^{\prime\prime}}$ are absent;
\item the $L^\prime$-conservation if $\mu_{\xi^\prime \eta k^\prime}$ is also absent.
\end{itemize}
Both the $L$- and $L^\prime$-conservations are not spontaneously broken.

Before discussing neutrino oscillations, we should mention another dangerous flavor-changing interactions of leptons due to the existence of $\rho$, $\rho^\prime$ and $\rho^{\prime\prime}$ because the charged leptons can simultaneously couple to these Higgs scalars. These interactions are found to be suppressed down to the phenomenologically acceptable level because the suppression factor is at most $m^2_\tau/v^2_{weak}$ from the incoming and outgoing vertices and provides enough suppression.  Since the approximate $L_e$ conservation is satisfied by our interactions, all $L_e$-changing flavor interactions such as $\mu \rightarrow e \gamma$ are more suppressed.  The detailed analyses have been described in Ref.\cite{ExtendedZeeMuTau}.  The existence of heavy leptons and extra gauge bosons as well as heavy exotic quarks \cite{331Related} may also disturb well-established weak-interaction phenomenology.  However, the additional contributions from these exotics will be well-suppressed because they are sufficiently heavy. Their masses controlled by the VEV's of $\chi$ and $\chi^{\prime}$ are taken to be $\sim 2$ TeV for the later analyses.

The heavy lepton mass matrix is simply given by the diagonal masses computed to be $m_{E^e}=f_E v_\chi$, $m_{E^+}=f^+_E v_\chi$ and $m_{E^-}=f^-_E v_{\chi^\prime}$. On the other hand, the charged lepton mass matrix has the following non-diagonal form:
\begin{eqnarray}
M_\ell= \left(
    \begin{array}{ccc}
    m_\ell^{ee} & 0                & 0                 \\
    0           & m_\ell^{\mu\mu}  & m_\ell^{\mu\tau}  \\
    0           & m_\ell^{\tau\mu} & m_\ell^{\tau\tau} \\
    \end{array}
    \right),
\label{Eq:Mell}
\end{eqnarray}
where
\begin{eqnarray}
m_\ell^{ee}&=&f_ev_\rho(=m_e),
\nonumber \\
m_\ell^{\mu\mu}&=&\frac{1}{2} 
    \left[ \left( f^+_\ell+ g^-_\ell\right) v_\rho - f^-_\ell v_{\rho^\prime} - g^+_\ell v_{\rho^{\prime\prime}}\right], \quad
m_\ell^{\mu\tau}=\frac{1}{2} 
    \left[ \left( f^+_\ell- g^-_\ell\right) v_\rho + f^-_\ell v_{\rho^\prime} - g^+_\ell v_{\rho^{\prime\prime}}\right], \quad
\nonumber \\
m_\ell^{\tau\mu}&=&\frac{1}{2} 
    \left[ \left( f^+_\ell - g^-_\ell\right) v_\rho - f^-_\ell v_{\rho^\prime} + g^+_\ell v_{\rho^{\prime\prime}}\right], \quad
m_\ell^{\tau\tau}=\frac{1}{2} 
    \left[ \left( f^+_\ell+ g^-_\ell\right) v_\rho + f^-_\ell v_{\rho^\prime} + g^+_\ell v_{\rho^{\prime\prime}}\right].
\label{Eq:MellElements}
\end{eqnarray}
The diagonalized charged lepton matrix, $M_\ell^{diag}$, are obtained by the use of two unitary matrices of $U$ and $V$ as shown in the Appendix.  For the given mixing angles of $\alpha$ and $\beta$ in Eq.(\ref{Eq:UellVell}), the mass parameters in Eq.(\ref{Eq:MellElements}) can be determined by Eq.(\ref{Eq:mell_diag_to_nondiag}) in terms of $m_{\mu,\tau}$.

\section{\label{sec:4}Neutrino masses and oscillations}
The neutrino mass matrix is given by $M^H_\nu$ and $M^C_\nu$:
\begin{eqnarray}
M_\nu= M^H_\nu+M^C_\nu,
\label{Eq:Mnu331}
\end{eqnarray}
where $M^H_\nu$ ($M^C_\nu$) arises from the two-loop radiative mechanism based on Fig.\ref{Fig:HeavyTwoLoops} (Fig.\ref{Fig:ChargedTwoLoops}) for the heavy-lepton (charged-lepton) exchanges.  The effective mass terms are expressed by
\begin{eqnarray}
&&
\left(
\psi_{+L}\rho\chi^\prime
\right)^2, \quad
\psi_{eL}\rho\chi^\prime
\cdot
\psi_{-L}\rho\chi^\prime,
\label{Eq:EffectiveHeavy}
\end{eqnarray}
from Fig.\ref{Fig:HeavyTwoLoops} and by
\begin{eqnarray}
&&
\left(
\psi_{-L}\rho+\psi_{+L}\rho^{\prime\prime}
\right)
\chi
\cdot
\left(
\psi_{-L}\rho+\psi_{+L}\rho^{\prime\prime}
\right)
\chi^\prime
+(\chi\leftrightarrow\chi^\prime), \quad
\psi_{eL}\rho\chi
\cdot
\left(
\psi_{-L}\rho+\psi_{+L}\rho^{\prime\prime}
\right)
\chi^\prime
+(\chi\leftrightarrow\chi^\prime), 
\label{Eq:EffectiveCharged}
\end{eqnarray}
from Fig.\ref{Fig:ChargedTwoLoops}, where the product of the three particles reads $abc$=$\epsilon^{\alpha\beta\gamma} a_\alpha b_\beta c_\gamma$.  After, $\chi$, $\chi^\prime$, $\rho$ and $\rho^{\prime\prime}$ acquire VEV's, where some of these contribute to masses of the heavy leptons and of the charged leptons, the heavy-lepton contributions supply main terms of neutrino masses, which are either $S_2$-symmetric or $S_2$-antisymmetric, while the charged-lepton contributions supply minor terms of neutrino masses, which are the admixtures of the $S_2$-symmetric and $S_2$-antisymmetric contributions.  These neutrino masses are expressed as:
\begin{eqnarray}
M^H_\nu= m^H_\nu\left(
    \begin{array}{ccc}
    0        & -r^H_\nu & r^H_\nu \\
    -r^H_\nu & 1 & 1 \\
    r^H_\nu & 1 & 1 \\
    \end{array}
    \right), \quad
M^C_\nu=\left(
    \begin{array}{ccc}
    0        		& \delta m^{e\mu}_\nu 	& \delta m^{e\tau}_\nu \\
    \delta m^{e\mu}_\nu	& \delta m^{\mu\mu}_\nu 		& \delta m^{\mu\tau}_\nu \\
    \delta m^{e\tau}_\nu	& \delta m^{\mu\tau}_\nu 		& \delta m^{\tau\tau}_\nu \\
    \end{array}
    \right) ,
\label{Eq:MnuMassMatrix0}
\end{eqnarray}
where $M^H_\nu$ and $r^H_\nu$ are calculated to be:
\begin{eqnarray}
&&m_\nu^H
=
-\frac{1}{2}f^2_\xi 
\left(
	\lambda_{\eta\rho\xi\rho}v_\rho^2
	+
	\lambda_{\eta\rho^\prime \xi\rho^{\prime\prime}}v_{\rho^\prime }v_{\rho^{\prime\prime}}
\right)
\mu_{\xi\eta k} m_{E^-}^2I_H\left(m^2_{\xi}, m^2_{E^-}, m^2_{k^{++}}\right), \quad
\nonumber\\
&&r^H_\nu = \sqrt{2}
\frac
{
	f_{\xi^\prime}
}
{
	f_{\xi}
}
\frac
{
	\mu_{\xi^{\prime}\eta k^\prime}
}
{
	\mu_{\xi\eta k}
}
\frac
{
	f^{+E}_{k^\prime}
}
{
	f^{-E}_k
}
\frac
{
	m^2_{E^+}
}
{
	m^2_{E^-}
}
\frac
{
	I_H\left(m^2_{\xi^\prime}, m^2_{E^+}, m^2_{k^{\prime ++}}\right)
}
{
	I_H\left(m^2_{\xi}, m^2_{E^-}, m^2_{k^{++}}\right)
}
\label{Eq:MnuMassFromHeavy}
\end{eqnarray}
with
\begin{eqnarray}
&&
I_H\left( m^2,m^2_E,m^2_k\right)
=
\frac
{J\left({m_{E^-}^2,m^2,m^2_E,m_{\eta^-}^2};m^2_k\right)-J\left({m^2_E,m^2,m_{E^-}^2,m_{\xi^+}^2};m^2_k\right)}
{m_{\eta^-}^2-m_{\xi^+}^2},
\label{Eq:I_H} \\
&&
J\left(
	m_1^2,m_2^2,m_3^2,m_4^2;m^2_k
\right)
=
\frac
{G\left
	(m_1^2,m_2^2\right)G\left(m_3^2,m_4^2
\right)}
{m_k^2},
\label{Eq:J}\\
&&
G\left(x,y\right)
=
\frac
{1}
{16\pi^2}
\frac
{x\ln\left(x/m^2_k\right)-y\ln\left(y/m^2_k\right)}
{x-y},
\label{Eq:G}
\end{eqnarray}
under the assumption that the masses of $k^{++}$ and $k^{\prime ++}$ are heavy enough to neglect other masses in the integration. The masses used here are denoted by $m^2_\phi$ for the corresponding particle of $\phi$.  The explicit form of $J$ is given by
\begin{eqnarray}
&&
J\left(a,b,c,d; e\right) = \int\frac{d^4 k}{\left( 2\pi\right)^4}\frac{d^4 q}{\left( 2\pi\right)^4}
\frac{1}
{
\left( k^2-a\right)
\left( k^2-b\right)
\left( q^2-c\right)
\left( q^2-d\right)
\left( \left( k-q\right)^2-e\right)
},
\label{Eq:Integral}
\end{eqnarray}
which can be evaluated via the following formula
\begin{eqnarray}\label{Eq:OneLoopIntegral}
&&\int \frac{d^4 q}{\left( {2\pi } \right)^{4} }\frac{1}{\left( q^2  - a\right)\left( q^2  - b \right)\left( q^2  - c  \right)} = - \frac{i}{16\pi ^2 }\left[ \frac{a \ln a }{\left( a-b\right)\left( a-c \right)} + \frac{b \ln b }{\left( b  - a  \right)\left( b  - c \right)} + \frac{c \ln c}{\left( c - a \right)\left( c  - b \right)} \right].
\end{eqnarray}

To evaluate $M^C_\nu$, it is convenient to express $-\mathcal{L}_Y$ of Eq.(\ref{Eq:Yukawa}) in terms of the neutrinos in weak interactions and the diagonalized charged leptons, $\psi_\ell$=($e_L$, $\mu_L$, $\tau_L$)$^T$, $\nu$=($\nu_{e L}$, $\nu_{\mu L}$, $\nu_{\tau L}$)$^T$ and $\ell$=($e_R$, $\mu_R$, $\tau_R$)$^T$, whose relevant terms are:
\begin{eqnarray}
\overline{\nu^c}U^T{\bf f}_\xi U\psi_\ell{\bar\xi}^+
+
\overline{\nu^c}U^T{\bf f}_{\xi^\prime}U\psi_\ell{\bar\xi}^{\prime+}
+
\overline{\ell^c}V^T{\bf f}_kV\ell k^{++}
+
\overline{\ell^c}V^T{\bf f}_{k^\prime}V\ell k^{\prime ++}
+
\overline{\psi_\ell}M_\ell^{diag}\ell+({\rm h.c.}),
\label{Eq:YukawaLepton}
\end{eqnarray}
where
\begin{eqnarray}
{\bf f}_\xi
= 
\frac{f_\xi}{2}
\left(
	\begin{array}{ccc}
		0 & 0  & 0  \\
   		0 & 0 & 1  \\
   		0 & -1 & 0  \\
	\end{array}
\right), \quad
{\bf f}_{\xi^\prime}
= 
\frac{f_{\xi^\prime}}{\sqrt{2}}
\left(
	\begin{array}{ccc}
   		0 & -1 & -1  \\
   		1 & 0 & 0  \\
   		1 & 0 & 0  \\
	\end{array}
 \right),\quad
{\bf f}_k=
\frac{f_k^{+\ell}}{2}
\left(
	\begin{array}{ccc}
   		0 & 0 & 0  \\
   		0 & 1 & 1  \\
   		0 & 1 & 1  \\
	\end{array} 
\right), \quad
{\bf f}_{k^\prime}=
\frac{f_{k^\prime}^{+\ell}}{2}
\left(
	\begin{array}{ccc}
   		0 & 0 & 0  \\
   		0 & 1 & -1  \\
   		0 & -1 & 1  \\
	\end{array} 
\right).
\label{Eq:CouplingMatrixLepton}
\end{eqnarray}
It is not difficult to reach an effective lagrangian corresponding to Fig.\ref{Fig:ChargedTwoLoops}:
\begin{eqnarray}
{\mathcal L}_{eff}
&=&
\lambda_{\eta \chi \xi \chi^{\prime}} v_{\chi} v_{\chi^{\prime}} 
\sum\limits_{(\varphi,K) = (\xi,k) ,(\xi^{\prime},k^\prime)} \mu _{\varphi \eta K} 
\sum\limits_{m,m^\prime,n,n^\prime = e ,\mu,\tau}
\overline {\nu _{m}^c } 
\left(
	{U^T {\bf f}_\varphi ^T U} 
\right)_{mm^{\prime}} 
\left( 
	M^\ell_{diag}
\right)_{m^{\prime}m^{\prime}} 
\nonumber\\
&&\times 
\left(
	V^T {\bf f}_K V
\right)_{m^{\prime}n^{\prime}} 
I_\ell\left(
	m^2_{\bar \varphi^+},m^2_{K^{++}} 
\right)_{m^{\prime}n^{\prime} } 
\left( 
	M^{\ell}_{diag} 
\right)_{n^{\prime}n^{\prime}}
\left(
	U^T {\bf f}_\xi  U
\right)_{n^{\prime}n}
\nu _{n},
\label{Eq:EffectiveNuCharged}
\end{eqnarray}
where
\begin{eqnarray}
I_\ell\left(m_{\bar \varphi^+}^2,m_K^2 \right)_{ab} 
= 
\frac
{
	J\left(m_a^2,m_{\bar \varphi^+}^2 ,m_b^2 ,m_{\eta^-}^2;m^2_K \right) - J\left(m_a^2 ,m_{\bar \varphi^+}^2 ,m_b^2 ,m_{{\bar \xi}^+}^2;m^2_K \right)
}
{
	m_{\eta^-}^2-m_{{\bar \xi}^+}^2
}.
\label{Eq:I_ell}
\end{eqnarray}
Eq.(\ref{Eq:EffectiveNuCharged}) can be further calculated to be:
\begin{eqnarray}
{\mathcal L}_{eff}
&=&
\frac{1}{16}f^2_\xi f^{+\ell}_k
\lambda_{\eta \chi \xi \chi^{\prime}} v_{\chi} v_{\chi^{\prime}} 
\mu_{\xi \eta k} I_\ell\left(m^2_{{\bar \xi}^+},m^2_{k^{++}}\right)
\overline{\nu^c } 
\left(
\begin{array}{ccc}
0 & 0 & 0  \\
0 & m^2_\tau \left(c_\beta+s_\beta\right)^2 & -m_\mu m_\tau \left(c^2_\beta-s^2_\beta\right)  \\
0 & -m_\mu m_\tau \left(c^2_\beta-s^2_\beta\right) & m^2_\mu \left(c_\beta-s_\beta\right)^2  \\
\end{array} 
\right)
\nu
\nonumber \\
&-&
\frac{1}{16\sqrt{2}}f_\xi f_{\xi^\prime} f^{-\ell}_{k^\prime}
\lambda_{\eta \chi \xi \chi^{\prime}} v_{\chi} v_{\chi^{\prime}} 
\mu_{\xi^\prime \eta k^\prime} I_\ell\left(m^2_{\bar \xi^{\prime +}},m^2_{k^{\prime ++}}\right)
\overline{\nu^c } 
\left(
\begin{array}{ccc}
0 & X & Y  \\
X & 0 & 0  \\
Y & 0 & 0  \\
\end{array} 
\right)
\nu,
\label{Eq:NuMassChargedComputed}
\end{eqnarray}
where the integral of $I_\ell$ is defined as $I_\ell\left( m_1^2,m_2^2\right)_{ab} = \delta_{ab} I_\ell\left( m_1^2,m_2^2\right)$ with $m_{e,\mu,\tau}^2$ safely neglected in the denominators in the integral, and
\begin{eqnarray}
&&X=m^2_\tau \left( c_\alpha + s_\alpha\right)\left( c_\beta - s_\beta\right)^2
-m_\mu m_\tau\left( c_\alpha - s_\alpha\right)\left( c^2_\beta - s^2_\beta\right),
\nonumber\\
&&Y=m_\mu m_\tau \left( c_\alpha + s_\alpha\right)\left( c^2_\beta - s^2_\beta\right)
-m^2_\mu\left( c_\alpha - s_\alpha\right)\left( c_\beta + s_\beta\right)^2.
\label{Eq:NuMassCharged_ei_Entries}
\end{eqnarray}
Since
\begin{eqnarray}
{\mathcal L}_{eff}
&=&
-\frac{1}{2}\overline {\nu^c} M^C_\nu\nu,
\label{Eq:NuMassCharged}
\end{eqnarray}
the each entry of $\delta m_\nu$'s in Eq.(\ref{Eq:MnuMassMatrix0}) can be easily read off from Eq.(\ref{Eq:NuMassChargedComputed}).

To see the order of magnitudes of neutrino masses, we use the simplest choice of masses
\begin{eqnarray}
m^2_{e,\mu,\tau} \ll m^2_{\eta^-} \ll m^2_{E^-,E^+} \ll m^2_{\xi^+,\xi^{\prime +},{\bar \xi}^+,{\bar \xi}^{\prime +}}(=m^2) \ll m^2_{k^{++}, k^{\prime ++}}.
\label{Eq:SimplestMass}
\end{eqnarray}
This choice reduces the integrals of $I_H$ and $I_\ell$ to
\begin{eqnarray}
I_H\left(m^2,m_E^2,m^2_k\right)
=
-\left( \frac{1}{16\pi}\right)^2\frac{1}{m^2_km^2}\ln\frac{m^2_k}{m^2}\ln\frac{m^2}{m^2_E},
\quad
I_\ell\left( m^2,m^2_k \right)
=
-\left( \frac{1}{16\pi}\right)^2\frac{1}{m^2_km^2}\ln\frac{m^2_k}{m^2}\ln\frac{m^2}{m^2_{\eta^-}}.
\label{Eq:SimplestIntegral}
\end{eqnarray}
We also choose the following magnitudes of relevant parameters for neutrino masses: 
\begin{enumerate}
\item $v_\rho= v_{\rho^\prime} = v_{\rho^{\prime\prime}}= v_{weak}/20$ ($\sim$10 GeV) and $v_\eta = v_{weak}$, where $v_\rho$ that controls the $b$-quark mass can be safely set to be $\sim$10 GeV and $v_\eta$ that supplies the $t$-quark mass should be $\sim v_{weak}$, since $v_\eta$ and $v_{\rho,\rho^\prime,\rho^{\prime\prime}}$ are related to weak boson masses proportional to $v_{weak}^2 = \sum_{all} v_{Higgs}^2$.

\item $v_\chi = v_\chi^\prime = 10v_{weak}$ ($\sim$ 2 TeV) since $\chi$ and $\chi^\prime$ are the key fields for the symmetry breaking of $SU(3)_L \times U(1)_N \rightarrow SU(2)_L \times U(1)_Y$, leading to $v_{\chi,\chi^\prime} \gg v_{weak}$.

\item $m_{\eta^-}=v_{weak}$, $m_{E^\pm} =ev_\chi (\sim 0.5$ TeV), $m_{\xi^-}=m_{{\bar \xi}^-}=m_{\xi^{\prime -}}=m_{{\bar \xi}^{\prime -}}=v_\chi (\sim 2$ TeV) and $m_{k^{++},k^{\prime ++}} = v_\chi/e$ ($\sim$ 6 TeV) to satisfy Eq.(\ref{Eq:SimplestMass}). 

\item $\mu_{\xi\eta k}=ev_\chi$ with $\mu_{\xi^\prime\eta k^\prime}/\mu_{\xi\eta k}\sim 0.1$, which can be regarded as a natural relation \cite{tHooft} because the limit of $\mu_{\xi^\prime\eta k^\prime}\rightarrow$0 recovers the $L_e$-conservation.

\item $f_{\xi, \xi^\prime}$ = $f_k^{-E}$ = $f_{k^\prime}^{+E}$ = $-f_k^{+\ell}$ = $-f_{k^\prime}^{-\ell}$ = $e$ and $\lambda_{\eta\rho\xi\rho,\eta\rho^\prime\xi\rho^{\prime\prime},\eta\chi\xi\chi^\prime}\sim e$ so that higher loop effects are safely neglected.
\end{enumerate}
 This mass-setting provides sufficient suppression of exotic contributions in low-energy phenomenology.

To see the magnitude of parameters for neutrino oscillations, we omit the terms proportional to $m_\mu$ ($\ll m_\tau$) in Eq.(\ref{Eq:NuMassChargedComputed}) and obtain that
\begin{eqnarray}
&&m^H_\nu = 3.3\times 10^{-2}\frac{\lambda}{e}~{\rm eV}, \quad 
r^H_\nu = \sqrt{2}\frac{\mu_{\xi^\prime\eta k^\prime}}{\mu_{\xi\eta k}},
\label{Eq:MuMassRoughComputedHeavy}\\
&&
\delta m^{\mu\mu}_\nu = -6.6\times 10^{-3}\frac{\lambda_\ell}{e}\left( c_\beta+s_\beta\right)^2~{\rm eV}, \quad
\delta m^{e\mu}_\nu = \frac{6.6\times 10^{-3}}{\sqrt{2}}\frac{\lambda_\ell}{e} \frac{\mu_{\xi^{\prime}\eta k^\prime}}{\mu_{\xi\eta k}}
\left( c_\beta-s_\beta\right)^2\left( c_\alpha+s_\alpha\right)~{\rm eV},
\label{Eq:MuMassRoughComputedCharged}
\end{eqnarray}
where $\lambda=\lambda_{\eta\rho\xi\rho}=\lambda_{\eta\rho^\prime\xi\rho^{\prime\prime}}$ and $\lambda_\ell =\lambda_{\eta\chi\xi\chi^\prime}$.  By choosing the remaining parameters to be:
\begin{eqnarray}
&&
\frac{\lambda}{e} = 0.8, \quad
\frac{\lambda_\ell}{\lambda} = 2.0, \quad
\frac{\mu_{\xi^{\prime} \eta k^\prime}}{\mu_{\xi\eta k}} = 0.12,\quad
\label{Eq:RemaingParameters}
\end{eqnarray}
 so as to recover the observed data, we find using the estimates of Eqs.(\ref{Eq:MuMassRoughComputedHeavy}) and (\ref{Eq:MuMassRoughComputedCharged}) that, for the simplest case of $s_{\alpha,\beta}=0$,
\begin{eqnarray}
&&
\Delta m^2_{atm} = 2.2\times 10^{-3}~{\rm eV}^2,\quad 
\sin^22\theta_{atm} = 0.96,
\nonumber\\
&&
\Delta m^2_\odot = 7.3\times 10^{-5}~{\rm eV}^2,\quad 
\sin^22\theta_\odot = 0.79~(\tan^2\theta_\odot = 0.37),
\nonumber\\
&&
\sin\theta_{13} = 2.5\times 10^{-2},
\label{Eq:OurPrediction}
\end{eqnarray}
where $\cos2\theta_{atm} > 0$ (corresponding to $\Delta > 0$) is chosen, and
\begin{eqnarray}
&&
m_1= 3.4\times 10^{-3}~{\rm eV}, \quad
m_2= 9.2\times 10^{-3}~{\rm eV}, \quad
m_3= 4.8\times 10^{-2}~{\rm eV}.
\label{Eq:OurPredictionMass}
\end{eqnarray}
The mixing angles are determined by $\sin^22\theta_{atm}=1-4\Delta^2$ and $\sin^22\theta_\odot = 8/(8+x^2)$ with
\begin{eqnarray}
&&
\Delta = -\frac{\delta m^{\mu\mu}_\nu}{4m^H_\nu}, \quad
x = \sqrt{2}\frac{\lambda_2}{A},
\label{Eq:FixingParameters}
\end{eqnarray}
as in Eq(\ref{Eq:Angles}).  The requirement of $\sin^2\theta_\odot\sim 0.8$ (corresponding to $x\sim\sqrt{2}$) results in $\lambda_2\sim A$.

To see the dependence of the charged-lepton mixing angles, $\alpha$ and $\beta$, in neutrino masses and mixings, we perform numerical estimation by keeping the terms proportional to $m_\mu$ in Eq.(\ref{Eq:NuMassChargedComputed}) and show the results in Fig.\ref{Fig:NumericalEstimationAtm} - Fig.\ref{Fig:NumericalEstimationMass}.  It can be stated that 
\begin{enumerate}
\item As in Fig.\ref{Fig:NumericalEstimationAtm}, since the atmospheric neutrino mixing is dominated by the heavy-lepton contributions, the predictions of $\sin^22\theta_{atm}$ and $\Delta m^2_{atm}$ are less-dependent on the charged lepton contributions, where their $\alpha$-dependence hardly manifests itself in this graph because it reflects $1\leq \left(c_\alpha+s_\alpha\right)\leq\sqrt{2}$ in $\delta m^{e\mu}_\nu$ as in Eq.(\ref{Eq:MuMassRoughComputedCharged}), and similarly $m_3$ is also less-dependent on the charged lepton contributions as in Fig.\ref{Fig:NumericalEstimationMass}; 
\item As in Fig.\ref{Fig:NumericalEstimationSolar}, since the solar neutrino mixing heavily depends on the charged lepton contributions, the predictions of $\sin^22\theta_\odot$ and $\Delta m^2_\odot$ permitting the observed values vary with the magnitudes of $\lambda_\ell$ and charged-lepton mixing angles;
\item As in Fig.\ref{Fig:NumericalEstimation13}, since the $\sin\theta_{13}$=0 from the heavy lepton contributions, its magnitude remain suppressed to be $\sin\theta_{13}\mapleq$0.03 induced by the charged lepton contributions.
\end{enumerate}
From Fig.\ref{Fig:NumericalEstimationSolar}, we find that the effects of $m_\mu\neq 0$ in Eq.(\ref{Eq:OurPrediction}) lower (raise) the magnitude of $\sin^22\theta_\odot$ ($\Delta m^2_\odot$) from 0.79 ($7.3\times 10^{-5}$ eV$^2$) to 0.76 ($8.3\times 10^{-5}$ eV$^2$).  We note that, in the simple-minded case of $\lambda_\ell / \lambda = 1$ instead of $\lambda_\ell / \lambda = 2$ as in Eq.(\ref{Eq:RemaingParameters}), the neutrino oscillations are characterized by
\begin{eqnarray}
&&
\Delta m^2_{atm} = 2.3\times 10^{-3}~{\rm eV}^2,\quad 
\sin^22\theta_{atm} = 0.97,
\nonumber\\
&&
\Delta m^2_\odot = 7.0\times 10^{-5}~{\rm eV}^2,\quad 
\sin^22\theta_\odot = 0.85~(\tan^2\theta_\odot = 0.44),
\nonumber\\
&&
\sin\theta_{13} = \left( 1.3-1.4\right)\times 10^{-2},
\label{Eq:OurPrediction0}
\end{eqnarray}
for the mixing angles of the charged leptons at $\sin 2\beta$ = 0.75 with all values of $\sin \alpha (\geq 0)$ allowed.

\section{\label{sec:5}Summary}

We have discussed how to theoretically understand the observed properties of neutrino oscillations, which can be summarized as:
\begin{enumerate}
\item The atmospheric neutrino mixing is almost maximal and is characterized by $\sin^22\theta_{atm}\sim 1$;
\item The solar neutrino mixing deviates from the maximal mixing and points to $\sin^22\theta_\odot\sim 0.8$;
\item The $U_{e3}$-mixing is suppressed;
\item The squared mass differences show the hierarchy of $\Delta m^2_{atm}\gg\Delta m^2_\odot$.
\end{enumerate}
Our answers consist of
\begin{enumerate}
\item The atmospheric neutrino mixing is almost maximal because the dominant contributions to the neutrino mass matrix give
\begin{equation}
M_\nu =    
    \left(
    \begin{array}{ccc}
         0 & b   & -b \\
         b & d   & d \\
         -b & d   & d\\
    \end{array}
    \right),
	\label{Eq:OurMnuSummary}
\end{equation}
as in Eq.(\ref{Eq:OurMnu}), which also give $\sin\theta_{13}=0$, for $c_{23}>0$ and $s_{23}>0$ defined in Eq.(\ref{Eq:MNS}).
\item The solar neutrino mixing deviates from the maximal mixing because less-dominant contributions perturb Eq.(\ref{Eq:OurMnuSummary}), yielding $\sin^22\theta_\odot\sim 0.8$;
\item The $U_{e3}$-mixing is suppressed because the dominant contributions ensure the appearance of $U_{e3}$=0 and less-dominant contributions yield the suppressed $U_{e3}$;
\item The squared mass differences show the hierarchy of $\Delta m^2_{atm}\gg\Delta m^2_\odot$ because the tiny breaking of the electron number conservation allows the appearance of the suppressed $\Delta m^2_\odot$.
\end{enumerate}
It should be noted that the following neutrino mass texture instead of Eq.(\ref{Eq:OurMnuSummary}):
\begin{equation}
M^\prime_\nu =    
    \left(
    \begin{array}{ccc}
         0 & b   & b \\
         b & d   & d \\
         b & d   & d\\
    \end{array}
    \right),
	\label{Eq:OurMnuSummary2}
\end{equation}
yields $\sin\theta_{13}(\sim b/\sqrt{2}d)\neq 0$.  These observations are quite different from those based on the two-zero texture of neutrino masses \cite{TwoTextureZero}.

To realize the above scenario, we have adopted the model based on $SU(3)_L\times U(1)_N$ and have demonstrated that 
\begin{enumerate}
\item The dominant contributions arise from the two-loop radiative effects of the heavy leptons;
\item The less-dominant contributions arise from the two-loop radiative effects of the ordinary charged leptons.
\end{enumerate}
The model contains the specific Higgs scalars denoted by $\xi$=($\xi^{++}$, $\xi^+$, ${\bar \xi}^+$)$^T$ with $L_e$=0 and by $\xi^\prime$=($\xi^{\prime ++}$, $\xi^{\prime +}$, ${\bar \xi}^{\prime +}$)$^T$ with $L_e=-1$, which initiate the two-loop radiative mechanism.  These Higgs scalars can also initiate the one-loop radiative mechanism, which turns out to be forbidden by the discrete $Z_8$ symmetry that simultaneously forbids the appearance of dangerous flavor-changing interactions due to the direct Higgs exchanges inherent to this model.  The dominant texture of Eq.(\ref{Eq:OurMnuSummary}) is subject to the $S_2$-symmetric Yukawa interactions of the heavy leptons involving their diagonalized mass terms. The less-dominant charged lepton contributions receive the $S_2$-breaking ones due to the existence of the $S_2$-breaking charged-lepton masses.  It is argued that, in Eq.(\ref{Eq:OurMnuSummary}), the $\xi$-contributions yield the $d$-term while the $\xi^\prime$-contributions yield the $b$-term, which identically vanishes in the exact $L_e$-conservation.

The $S_2$-permutation symmetry for $\mu$-$\tau$ can describe the gross feature of the observed atmospheric neutrino oscillations.  Since the existing charged leptons obviously do not respect the $S_2$-permutation symmetry, there should be other ingredients that respect the $S_2$-permutation symmetry for the neutrino mixings.  If we rely upon the seesaw mechanism instead of the radiative mechanism discussed in this article, it is expected that the type II seesaw mechanism \cite{type2seesaw} is arranged to respect the $S_2$-permutation symmetry while the ordinary seesaw mechanism \cite{Seesaw} influenced by the charged leptons breaks the $S_2$-conservation, which as in the present discussions provides the significant deviation of the solar neutrino mixing from the maximal one.

\appendix
\section{Charged Lepton Masses}
In this appendix, the diagonalization of the lepton mass matrix $M_\ell$ is demonstrated.  The diagonal masses are obtained after the transformation by $M_\ell^{diag} = diag.(m_e,m_\mu,m_\tau) = U_\ell^\dagger M_\ell V_\ell$, where unitary matrices $U_\ell$ and $V_\ell$ are given by 
\begin{eqnarray}
U_\ell = \left(
    \begin{array}{ccc}
    1 & 0 &  0 \\
    0 & c_\alpha & s_\alpha \\
    0 & -s_\alpha & c_\alpha \\
    \end{array}
    \right), 
\quad
V_\ell = \left(
    \begin{array}{ccc}
    1 & 0 &  0 \\
    0 & c_\beta & s_\beta \\
    0 & -s_\beta & c_\beta \\
    \end{array}
    \right) 
\label{Eq:UellVell}
\end{eqnarray}
with $c_\alpha = \cos \alpha,$ etc., defined by
\begin{eqnarray}
c_\alpha &=& \sqrt{\frac{(m_\ell^{\tau\tau})^2+(m_\ell^{\tau\mu})^2-m_\mu^2}{m_\tau^2-m_\mu^2}},
\quad
s_\alpha = \sqrt{\frac{-(m_\ell^{\tau\tau})^2-(m_\ell^{\tau\mu})^2+m_\tau^2}{m_\tau^2-m_\mu^2}},
\nonumber \\
c_\beta &=& \sqrt{\frac{(m_\ell^{\tau\tau})^2+(m_\ell^{\mu\tau})^2-m_\mu^2}{m_\tau^2-m_\mu^2}},
\quad
s_\beta = \sqrt{\frac{-(m_\ell^{\tau\tau})^2-(m_\ell^{\mu\tau})^2+m_\tau^2}{m_\tau^2-m_\mu^2}},
\label{Eq:calpha_salpha}
\end{eqnarray}
where $m^2_\mu+m^2_\tau$=$(m_\ell^{\mu\mu})^2$+$(m_\ell^{\tau\tau})^2$+$(m_\ell^{\mu\tau})^2$+$(m_\ell^{\tau\mu})^2$ is a obvious relation. The diagonal masses are computed to be:
\begin{eqnarray}
m_e^2&=&(m_\ell^{ee})^2,
\nonumber \\
m_\mu^2&=&\frac{1}{2}
    \left[ (m_\ell^{\tau\tau})^2+(m_\ell^{\mu\mu})^2+(m_\ell^{\tau\mu})^2+(m_\ell^{\mu\tau})^2 - M^2\right],
\nonumber \\
m_\tau^2&=&\frac{1}{2}
    \left[ (m_\ell^{\tau\tau})^2+(m_\ell^{\mu\mu})^2+(m_\ell^{\tau\mu})^2+(m_\ell^{\mu\tau})^2 + M^2\right],
\label{Eq:mell_diagonalmass}
\end{eqnarray}
where
\begin{eqnarray}
M^4 &=& \left[ (m_\ell^{\tau\tau})^2-(m_\ell^{\mu\mu} )^2\right]^2
      + \left[ (m_\ell^{\tau\mu} )^2-(m_\ell^{\mu\tau})^2\right]^2
\nonumber \\ 
	&&+ 2(m_\ell^{\tau\tau}m_\ell^{\mu\tau}+m_\ell^{\mu\mu}m_\ell^{\tau\mu})^2
	  + 2(m_\ell^{\tau\tau}m_\ell^{\tau\mu}+m_\ell^{\mu\mu}m_\ell^{\mu\tau})^2.
\label{Eq:M4}
\end{eqnarray}
There are the following relations for non-diagonal and diagonal charged lepton masses:
\begin{eqnarray}
m_\ell^{\mu\mu}&=&S^2m_\tau + C^2m_\mu, \quad m_\ell^{\tau\tau}=C^2m_\tau + S^2m_\mu,
\nonumber \\
m_\ell^{\mu\tau}&=&\frac{1}{c_\beta^2-s_\alpha^2}
    \left[ (c_\alpha s_\alpha C^2 - c_\beta  s_\beta  S^2)m_\tau \right.
   -\left. (c_\beta  s_\beta  C^2 - c_\alpha s_\alpha S^2)m_\mu  \right],
\nonumber \\
m_\ell^{\tau\mu}&=&\frac{1}{c_\beta^2-s_\alpha^2}
    \left[ (c_\beta  s_\beta  C^2 - c_\alpha s_\alpha S^2)m_\tau \right.
   -\left. (c_\alpha s_\alpha C^2 - c_\beta  s_\beta  S^2)m_\mu  \right],
\label{Eq:mell_diag_to_nondiag}
\end{eqnarray}
where $C^2$ and $S^2$ are defined by 
\begin{eqnarray}
C^2=\frac{c_\alpha^2+c_\beta^2}{2}, \quad
S^2=\frac{s_\alpha^2+s_\beta^2}{2}.
\label{Eq:C2S2}
\end{eqnarray}


\noindent

\begin{center}
\textbf{Table Captions}
\end{center}
\begin{description}
\item{TABLE \ref{Tab:particlesAndSymmetries}:} 
Particle contents with $S_{L2}$, $Z_8$ and $L_e$ quantum numbers for quarks, leptons and Higgs scalars, where $S_{2L}=+(-)$ denotes symmetric (antisymmetric) states and $\omega^8$=1. 
\end{description}

\bigskip
\noindent

\begin{figure}[!htbp]
\begin{flushleft}
\includegraphics*[3mm,1mm][200mm,70mm]{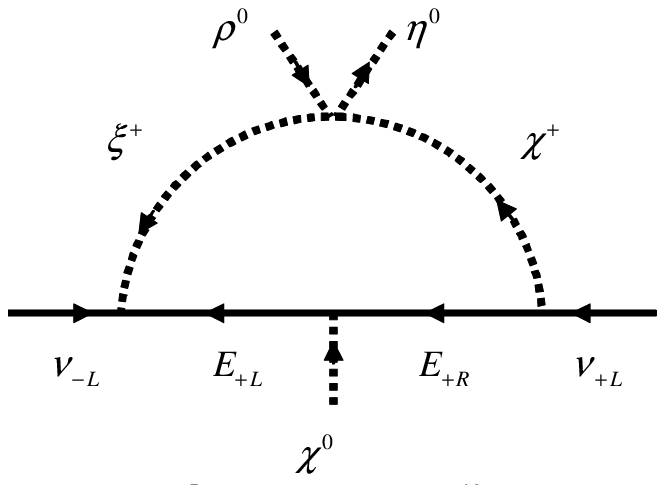}
\end{flushleft}
\caption{Forbidden one-lopp diagram.}
\label{Fig:WouldBeOneLoop}
\end{figure}

\begin{figure}[!htbp]
 \includegraphics*[3mm,1mm][200mm,70mm]{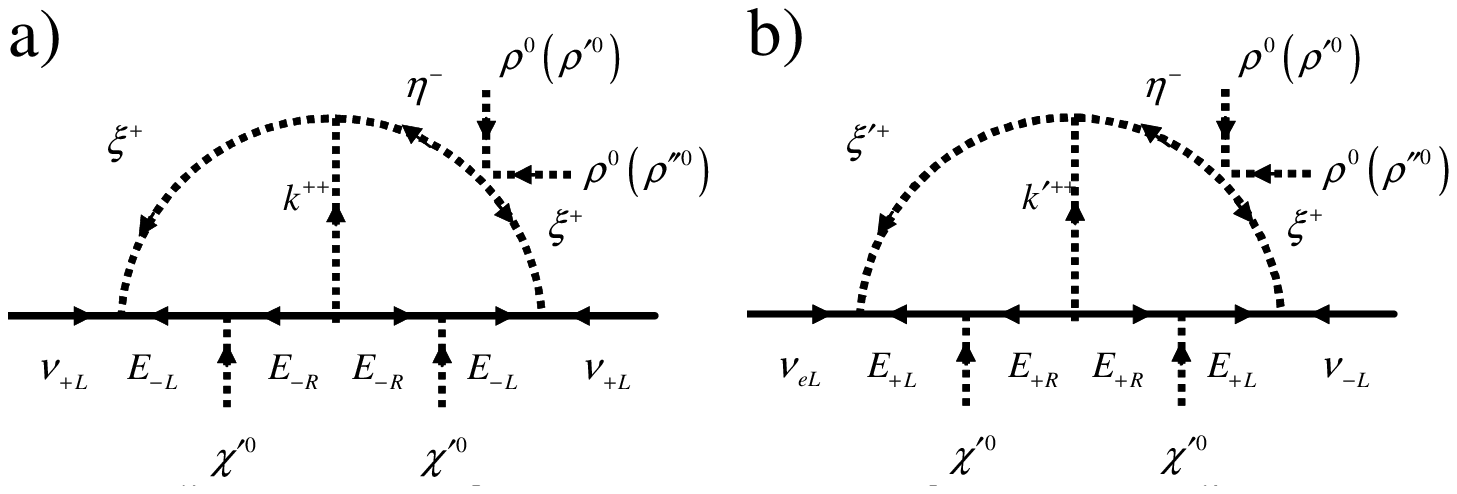}
  \caption{Heavy-lepton mediated two-loop diagrams.}
\label{Fig:HeavyTwoLoops}
\end{figure}

\begin{figure}[!htbp]
\includegraphics*[3mm,1mm][200mm,70mm]{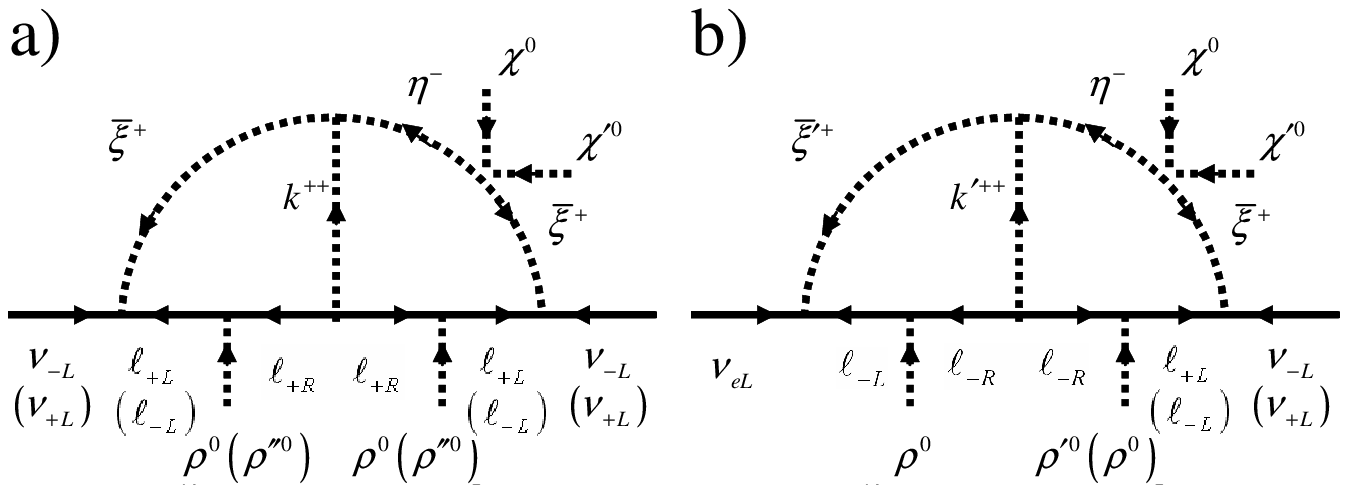}
  \caption{Charged-lepton mediated two-loop diagrams.}
\label{Fig:ChargedTwoLoops}
\end{figure}

\begin{figure}[!htbp]
\begin{flushleft}
\includegraphics*[3mm,1mm][200mm,152mm]{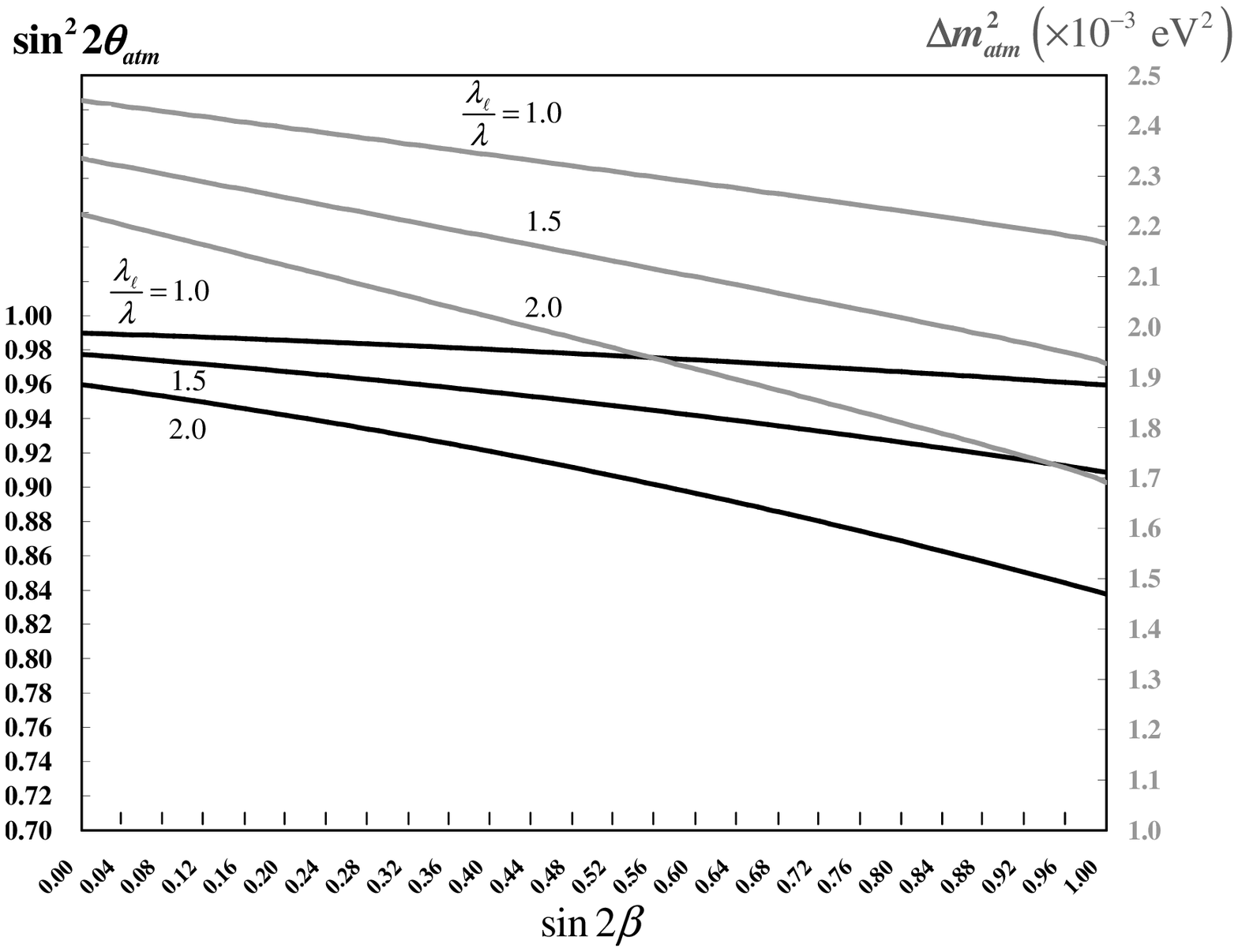}
\end{flushleft}
  \caption{The dependences of $\sin^22\theta_{atm}$ (the black curves) and $\Delta m^2_{atm}$ (the gray curves) in functions of $\sin 2\beta$ for the given values of $\lambda_\ell/\lambda$=$1.0, 1.5, 2.0$, where the dependence in $\sin\alpha$ is so weak that it is covered by the line thickness.}
\label{Fig:NumericalEstimationAtm}
\end{figure}

\begin{figure}[!htbp]
\begin{flushleft}
\includegraphics*[3mm,1mm][200mm,152mm]{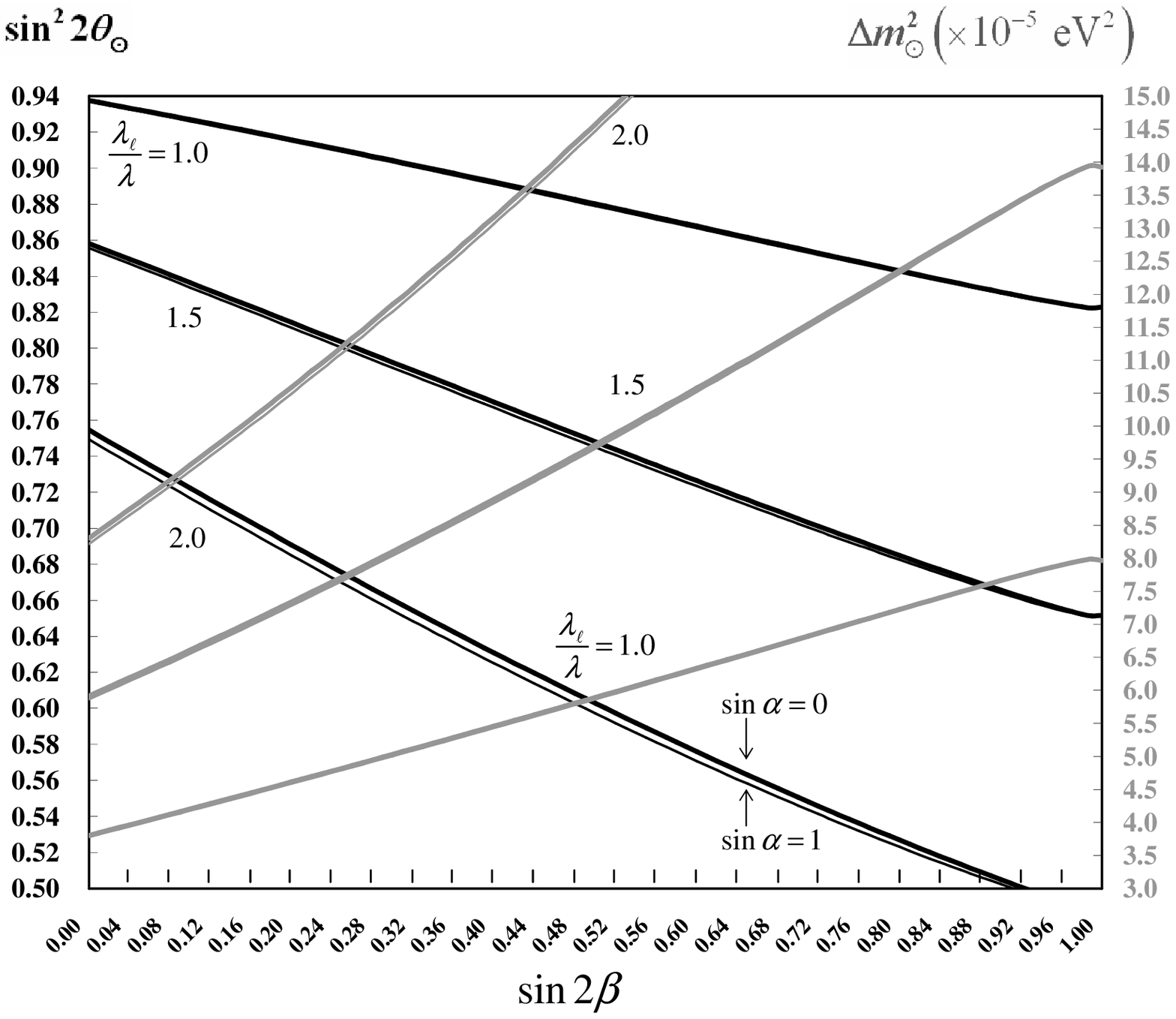}
\end{flushleft}
  \caption{The dependences of $\sin^22\theta_\odot$ (the black curves) and $\Delta m^2_\odot$ (the gray curves) in functions of $\sin 2\beta$ for the given values of $\sin\alpha$=0 (thick curves), 1 (thin curves) and $\lambda_\ell/\lambda$=$1.0, 1.5, 2.0$.}
\label{Fig:NumericalEstimationSolar}
\end{figure}

\begin{figure}[!htbp]
\begin{flushleft}
\includegraphics*[3mm,1mm][200mm,150mm]{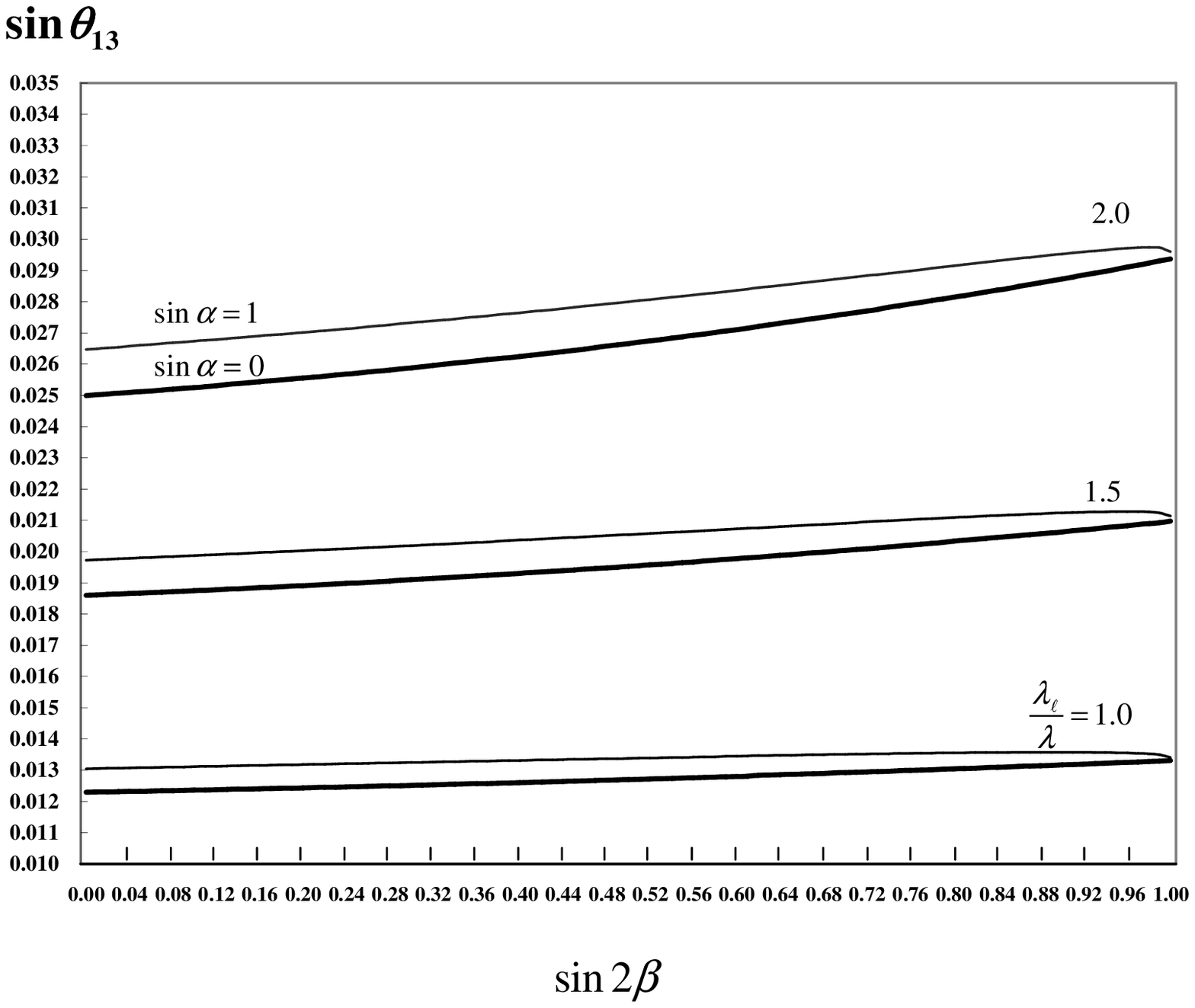}
\end{flushleft}
  \caption{The dependences of $\sin\theta_{13}$ as a function of $\sin 2\beta$ for the given values of $\sin\alpha$=0 (thick curves), 1 (thin curves) and $\lambda_\ell/\lambda$=$1.0, 1.5, 2.0$.}
\label{Fig:NumericalEstimation13}
\end{figure}

\begin{figure}[!htbp]
\begin{flushleft}
\includegraphics*[3mm,1mm][200mm,150mm]{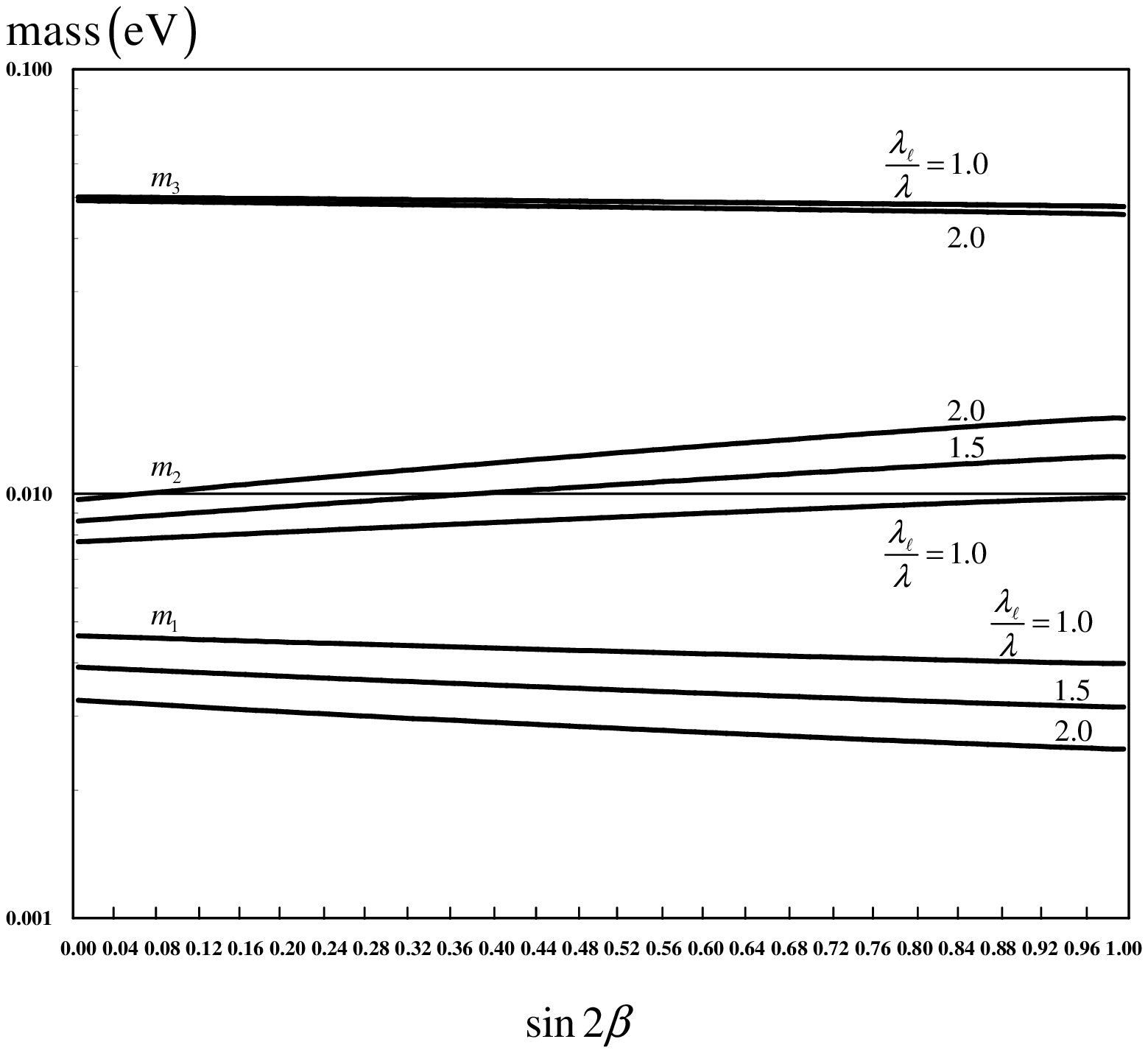}
\end{flushleft}
  \caption{The dependences of $m_{1,2,3}$ as a function of $\sin 2\beta$ for the given values of $\lambda_\ell/\lambda$=$1.0, 1.5, 2.0$, where the dependence in $\sin\alpha$ is so weak that it is covered by the line thickness.}
\label{Fig:NumericalEstimationMass}
\end{figure}

\begin{center}
\textbf{Figure Captions}
\end{center}
\begin{description}
\item{FIG.\ref{Fig:WouldBeOneLoop}:}
Forbidden one-lopp diagram.
\item{FIG.\ref{Fig:HeavyTwoLoops}:}
Heavy-lepton mediated two-loop diagrams. 
\item{FIG.\ref{Fig:ChargedTwoLoops}:}
Charged-lepton mediated two-loop diagrams. 	
\item{FIG.\ref{Fig:NumericalEstimationAtm}:}
The dependences of $\sin^22\theta_{atm}$ (the black curves) and $\Delta m^2_{atm}$ (the gray curves) in functions of $\sin 2\beta$ for the given values of $\lambda_\ell/\lambda$=$1.0, 1.5, 2.0$, where the dependence in $\sin\alpha$ is so weak that it is covered by the line thickness.
\item{FIG.\ref{Fig:NumericalEstimationSolar}:}
The dependences of $\sin^22\theta_\odot$ (the black curves) and $\Delta m^2_\odot$ (the gray curves) in functions of $\sin 2\beta$ for the given values of $\sin\alpha$=0 (thick curves), 1 (thin curves) and $\lambda_\ell/\lambda$=$1.0, 1.5, 2.0$.
\item{FIG.\ref{Fig:NumericalEstimation13}:}
The dependences of $\sin\theta_{13}$ as a function of $\sin 2\beta$ for the given values of $\sin\alpha$=0 (thick curves), 1 (thin curves) and $\lambda_\ell/\lambda$=$1.0, 1.5, 2.0$.
\item{FIG.\ref{Fig:NumericalEstimationMass}:}
The dependences of $m_{1,2,3}$ as a function of $\sin 2\beta$ for the given values of $\lambda_\ell/\lambda$=$1.0, 1.5, 2.0$, where the dependence in $\sin\alpha$ is so weak that it is covered by the line thickness.
\end{description}

\begin{table}[ht]
    \caption{Particle contents with $S_{L2}$, $Z_8$ and $L_e$ quantum numbers for quarks, leptons and Higgs scalars, where $S_{2L}=+(-)$ denotes symmetric (antisymmetric) states and $\omega^8$=1. 
	\label{Tab:particlesAndSymmetries}}
    \begin{center}
    \begin{tabular}{ccccccccccc}
    \hline
           & $\psi_L^e$ & $e_R$     & $\psi_{+ L}$ & $\ell_{+ R}$ & $\psi_{- L}$ & $\ell_{- R}$ & $E^e_R$ & $E_{+ R}$ & $E_{- R}$&\\
    \hline
$S_{2L}$   & $+$        & $+$       & $+$          & $+$          & $-$           & $-$         & $+$     & $+$       &  $-$ &\\
$Z_8$      & $\omega^6$ & $\omega^4$& $\omega^5$  & $\omega^3$   &  $1$          & $\omega^6$   & $\omega^3$& $\omega^2$&$\omega^7$&\\
    \hline
$L$        & $1$        & $1$        & $1$         &$1$           & $1$           & $1$          & $1$     & $1$      & $1$ &\\    
$L_e$      & $1$        & $1$        & $0$         &$0$           & $0$           & $0$          & $1$     & $0$      & $0$ &\\    
    \hline\hline
          & $\eta$     & $\rho$     &$\rho^\prime$& $\rho^{\prime\prime}$& $\chi$& $\chi^\prime$& $\xi$  & $\xi^\prime$& $k^{++}$ & $k^{\prime ++}$\\
    \hline
$S_{2L}$  & $+$        & $+$        & $-$         & $-$          & $+$           & $+$         & $-$     & $+$        &  $+$ &+\\
$Z_8$     & $\omega$   & $\omega^2$ & $\omega^7$  & $\omega^5$   & $\omega^3$    & $\omega$     & $\omega^3$& $\omega^5$&$\omega^2$&$\omega^4$\\
    \hline
$L$       & $0$        & $0$        & $0$         &$0$           & $0$           & $0$          & $-2$    & $-2$      & $-2$ &$-2$\\    
$L_e$     & $0$        & $0$        & $0$         &$0$           & $0$           & $0$          & $0$     & $-1$      & $0$ &0\\    
    \hline
    \end{tabular}
    \end{center}
\end{table}

\noindent

\begin{thebibliography}{}
\bibitem{K2K} 
	S.H. Ahn {\it et al}. [K2K Collaboration], \Journal{\PLB}{511}{178}{2001}; \Journal{\PRL}{90}{041801}{2003};
	T. Ishii,  talk at {\it NOON2004: The 5th Workshop on Neutrino Oscillations and their Origin}, Tokyo, Japan (Feb. 11-15, 2004).

\bibitem{KamLAND} 
	K. Eguchi {\it et al}. [KamLAND collaboration], \Journal{\PRL}{90}{021802}{2003};
	S. Freedman,  talk at {\it NOON2004: The 5th Workshop on Neutrino Oscillations and their Origin}, Tokyo, Japan (Feb. 11-15, 2004).

\bibitem{Kamiokande}
	Y. Fukuda {\it et al}. [Super-Kamiokande Collaboration], \Journal{\PRL}{81}{1158}{1998} 
	[\Journal{\Erratum}{81}{4297}{1998}]; \Journal{\PRL}{82}{2430}{1999}.
	S. Fukuda {\it et al}. [Super-Kamiokande Collaboration], \Journal{\PRL}{86}{5651,5656}{2001}; 
	\Journal{\PLB}{539}{179}{2002};
	Y. Takeuchi, \Journal{\IJMP}{18}{3777}{2003};
	S. Koshio,  talk at {\it NOON2004: The 5th Workshop on Neutrino Oscillations and their Origin}, Tokyo, Japan (Feb. 11-15, 2004);
	C. Saji, talk at {\it NOON2004: The 5th Workshop on Neutrino Oscillations and their Origin}, Tokyo, Japan (Feb. 11-15, 2004). 

\bibitem{SNO}
	Q.R. Ahmad {\it et al}. [SNO Collaboration], \Journal{\PRL}{87}{071301}{2001}; \Journal{\PRL}{89}{011301}{2002};
	S. Graham,  talk at {\it NOON2004: The 5th Workshop on Neutrino Oscillations and their Origin}, Tokyo, Japan (Feb. 11-15, 2004).
\bibitem{RecentAnalyses}
	See for example, 
	M.C. Gonzalez-Garcia, \Journal{\RMP}{75}{345}{2003}; 
	talk at {\it NOON2004: The 5th Workshop on Neutrino Oscillations and their Origin}, Tokyo, Japan (Feb. 11-15, 2004);
	V. Barger and D. Marfatia, \Journal{\PLB}{555}{144}{2003};
	V. Barger, D. Marfatia and K. Whisnant, \Journal{\IJMPE}{12}{569}{2003}.
\bibitem{MassiveNeutrino} 
    Z. Maki, M. Nakagawa and S. Sakata, \Journal{\PTP}{28}{870}{1962}. 
    See also  B. Pontecorvo, \Journal{\JETPUSSR}{34}{247}{1958};  
    B. Pontecorvo, \Journal{\ZETP}{53}{1717}{1967};
    V. Gribov and B. Pontecorvo, \Journal{\PLBOLD}{28B}{493}{1969}.
\bibitem{Bimaximal}
	F. Vissani, arXiv:hep-ph/9708483;
    D. V. Ahluwalia, \Journal{\MPL}{13}{2249}{1998};
    V. Barger, P. Pakvasa, T.J. Weiler and K. Whisnant, \Journal{\PLB}{437}{107}{1998};
    A.J. Baltz, A.S. Goldhaber and M. Goldhaber, \Journal{\PRL}{81}{5730}{1998};
    M. Jezabek and Y. Sumino, \Journal{\PLB}{440}{327}{1998};
    R.N. Mohapatra and S. Nussinov, \Journal{\PLB}{441}{299}{1998};
    Y. Nomura and T. Yanagida, \Journal{\PRD}{59}{017303}{1999};
    I. Starcu and D.V.Ahluwalia, \Journal{\PLB}{460}{431}{1999};
    Q. Shafi and Z. Tavartkiladze, \Journal{\PLB}{451}{129}{1999};\Journal{\PLB}{482}{145}{2000};
    C.H. Albright and S.M. Barr, \Journal{\PLB}{461}{218}{1999};
    H. Georgi and S.L. Glashow, \Journal{\PRD}{61}{097301}{2000};
    R.N. Mohapatra, A. P\'{e}rez-Lorenzana and C. A. de S. Pires, \Journal{\PLB}{474}{355}{2000};
	B. Brahmacari and S. Choubey, \Journal{\PLB}{531}{99}{2002}
	K. S. Babu and R. N. Mohapatra, \Journal{\PLB}{532}{77}{2002};
	R. Kuchimanchi and R. N. Mohapatra, \Journal{\PRD}{66}{051301}{2002}.
	C. Giunti and M. Tanimoto, \Journal{\PRD}{66}{053013}{2002};
\bibitem{NearlyBimaximal} 
    H. Fritzsch and Z.Z. Xing, \Journal{\PLB}{372}{265}{1996}; \Journal{\PLB}{440}{313}{1998};
    M. Fukugita, M. Tanimoto and T. Yanagida, \Journal{\PRD}{57}{4429}{1998}; 
    M. Tanimoto, \Journal{\PRD}{59}{017304}{1999}. 
\bibitem{Seesaw} 
    T. Yanagida, in {\it Proceedings of the Workshop on Unified Theories and 
    Baryon Number in the Universe} edited by A. Sawada and A. Sugamoto 
    (KEK Report No.79-18, Tsukuba, 1979), p.95; \Journal{\PTP}{64}{1103}{1980};  
    M. Gell-Mann, P. Ramond and R. Slansky, in {\it Supergravity} edited by P. van 
    Nieuwenhuizen and D.Z. Freedmann (North-Holland, Amsterdam 1979), p.315; 
    R.N. Mohapatra and G. Senjanovi\'{c}, \Journal{\PRL}{44}{912}{1980}.
\bibitem{type2seesaw}
    R.N. Mohapatra and G. Senjanovi\'{c}, \Journal{\PRD}{23}{165}{1981};
    C. Wetterich, \Journal{\NPB}{187}{343}{1981}.
\bibitem{Zee}
    A. Zee, \Journal{\PLBOLD}{93B}{389}{1980}, in Ref.\cite{Lprime};
    L. Wolfenstein, \Journal{\NPB}{175}{93}{1980},
	S. T. Petcov, \Journal{\PLBOLD}{115B}{401}{1982}.
\bibitem{Babu}
    A. Zee, \Journal{\NPBOLD}{264B}{99}{1986}; 
    K. S. Babu, \Journal{\PLB}{203}{132}{1988}; 
    D. Chang, W-Y.Keung and P.B. Pal, \Journal{\PRL}{61}{2420}{1988}; 
    J. Schechter and J.W.F. Valle, \Journal{\PLB}{286}{321}{1992}.
\bibitem{Lprime}
    R. Barbieri, L.J. Hall, D. Smith, N.J. Weiner and A. Strumia, \Journal{\JHEP}{12}{017}{1998}.
	See also
    S.T. Petcov, \Journal{\PLBOLD}{110B}{245}{1982};
    C.N. Leung and S.T. Petcov, \Journal{\PLBOLD}{125B}{461}{1983};
    A. Zee, \Journal{\PLBOLD}{161B}{141}{1985}.
\bibitem{ExtendedZeeMuTau}
    T. Kitabayashi and M. Yasu\`{e}, \Journal{\PLB}{524}{308}{2002}; 
	\Journal{\IJMP}{17}{2519}{2002}.
\bibitem{MuTauYasu}
	T. Kitabayashi and M. Yasu\`{e}, \Journal{\PRD}{67}{015006}{2003}. 
\bibitem{GrimusMuTau}
	W. Grimus and L. Lavoura, \Journal{\JHEP}{0107}{045}{2001}; \Journal{\EPJ}{28}{123}{2003}; \Journal{\PLB}{573}{189}{2003};\Journal{\JPG}{30}{73}{2004};
	W. Grimus, talk at {\it NOON2004: The 5th Workshop on Neutrino Oscillations and their Origin}, Tokyo, Japan (Feb. 11-15, 2004). 
\bibitem{MuTau}
    E. Ma, arXiv:hep-ph/0208097;
	P.F. Harrison and W.G. Scott, \Journal{\PLB}{547}{219}{2002}. 
\bibitem{ZeeMaximal}
    C. Jarlskog, M. Matsuda, S. Skadhauge and M. Tanimoto, \Journal{\PLB}{449}{240}{1999};
	P. H. Frampton and S. L. Glashow, \Journal{\PLB}{461}{95}{1999};
	Y. Koide and A. Ghosal, \Journal{\PRD}{63}{037301}{2001};
	Y. Koide, \Journal{\PRD}{64}{077301}{2001};
	P. H. Frampton, M. C. Oh and T. Yoshikawa, \Journal{\PRD}{65}{073014}{2002};
	G. Dutta, arXiv:hep-ph/0202097 (Feb., 2002);
\bibitem{ExtendedZee}
	N. Gaur, A. Ghosal, E. Ma and P. Roy, \Journal{\PRD}{58}{071301}{1998};
    K.R.S. Balaji, W. Grimus and T. Schwetz, \Journal{\PLB}{508}{301}{2001}; 
	P. Roy and S. K. Vempati, \Journal{\PRD}{65}{073011}{2002};
	K.S. Babu and C. Macesanu. \Journal{\PRD}{67}{073010}{2003};
	K. Hasegawa, C.S. Lim and K. Ogure, \Journal{\PRD}{68}{053006}{2003};
	Xiao-Gang He, arXiv:hep-ph/0307172 
\bibitem{331}
    F. Pisano and V. Pleitez, \Journal{\PRD}{46}{410}{1992};
    P.H. Frampton, \Journal{\PRL}{69}{2889}{1992};
    D. Ng, \Journal{\PRD}{49}{4805}{1994};
	M. \"{O}zer, \Journal{\PRD}{54}{1143}{1996}.
	See also, M. Singer, J. W. F. Valle and J. Schechter, \Journal{\PRD}{22}{738}{1980}.
\bibitem{331RadiativeNu}
    J. W. F. Valle and M. Singer, \Journal{\PRD}{28}{540}{1983};
    R. Barbieri and R. N. Mohapatra, \Journal{\PLB}{218}{225}{1989};
    J. Liu, \Journal{\PLB}{225}{148}{1989};
    R. Foot, O. F. Hern\'{a}ndez, F. Pisano and V. Pleitez, \Journal{\PRD}{47}{4158}{1993};
    V. Pleitez and M. D. Tonasse, \Journal{\PRD}{48}{5274}{1993}; 
    P. H. Frampton, P.I. Krastev and J.T. Liu, \Journal{\MPL}{9}{761}{1994};
	F. Pisano, V. Pleitez and M. D. Tonasse, arXiv:hep-ph/9310230 v2 (Feb., 1994);
    M. B. Tully and G. C. Joshi, \Journal{\PRD}{64}{011301}{2001}; \Journal{\IJMP}{18}{1573}{2003}; 
	J. C. Montero, C. A. de S. Pires and V. Pleitez, \Journal{\PRD}{65}{093017}{2002};
	\Journal{\PRD}{65}{095001}{2002};
	J. C. Montero, V. Pleitez and M. C. Rodriguez, \Journal{\PRD}{65}{095008}{2002};
	A. Gusso, C. A. de S. Pires and P. S. Rodrigues da Silva, \Journal{\JPG}{30}{37}{2004}.
\bibitem{331Rad}
    Y. Okamoto and M. Yasu\`{e}, \Journal{\PLB}{466}{267}{1999};
    T. Kitabayashi and M. Yasu\`{e}, \Journal{\PRD}{63}{095002}{2001};
	\Journal{\PRD}{63}{095006}{2001}; \Journal{\NPB}{609}{61}{2001};
	T. Kitabayashi, \Journal{\PRD}{64}{057301}{2001}.
\bibitem{331HeavyERad}
	T. Kitabayashi and M. Yasu\`{e}, \Journal{\PLB}{508}{85}{2001}. 
\bibitem{331HeavyE}
	M. \"{O}zer, in Ref.\cite{331}.
\bibitem{Ue3}
    The Chooz Collaboration, M. Appollonio et al., \Journal{\PLB}{466}{415}{1999};
	The PALOVERDE Collaboration, F. Boehm et al., \Journal{\PRD}{62}{072002}{2000}.
\bibitem{MassMatrix}
   T. Fukuyama and H. Nishiura, in {\it Proceedings of International Workshop on Masses and Mixings of Quarks and Leptons} edited by Y. Koide (World Scientific, Singapore, 1997), p.252; hep-ph/9702253;
   E. Ma, M. Raidal and U. Sarkar, \Journal{\PRL}{85}{3769}{2000};
   S.K. Kang and C.S. Kim, \Journal{\PRD}{63}{113010}{2001};
   W. Grimus and L. Lavoura,  in Ref.\cite{GrimusMuTau}.
   See also, I. Dorsner and S.M. Barr, \Journal{\NPB}{617}{493}{2001}.
\bibitem{FCNCSU3}
    J.C. Montero, F. Pisano and V. Pleitez, \Journal{\PRD}{47}{2918}{1993};
    M. \"{O}zer, \Journal{\PRD}{54}{4561}{1996}.
\bibitem{FCNC}
    S.L. Glashow and S. Weinberg, \Journal{\PRD}{15}{1958}{1977};
    H. Georgi and A. Pais, \Journal{\PRD}{19}{2746}{1979}.
\bibitem{331Related}
    See for example,
	F-z. Chen, \Journal{\PLB}{442}{223}{1998};
    P.H. Frampton and M. Harada, \Journal{\PRD}{58}{095013}{1998};
	M.B. Tully and G.C. Joshi, \Journal{\IJMP}{13}{5593}{1998}; \Journal{\PLB}{466}{333}{1999};
	P.H. Frampton and A. Rasin, \Journal{\PLB}{482}{1293}{2000};
	N. A. Ky, H. N. Long and D. V. Soa, \Journal{\PLB}{486}{140}{2000};
	H. N. Long and D. V. Soa, \Journal{\NPB}{601}{361}{2001};
	E.M. Gregores, A. Gusso and S.F. Novaes, \Journal{\PRD}{64}{015004}{2001};
    H.N. Long and Le P. Trung, \Journal{\PLB}{502}{63}{2001}.
\bibitem{tHooft}
    G. 't Hooft, in: G. 't Hooft et al. (Eds.), Recent Development in Gauge Theories, 
	Proceedings of the Cargese Summer Institute, Cargese, France, 1979, NATO Adv. Sci. Inst. Ser. B Phys.,
	Vol. 59, Plenum Press, New York, 1980.
\bibitem{TwoTextureZero}
   P.H. Frampton, S.L. Glashow and D. Marfatia, \Journal{\PLB}{536}{79}{2002}.
   See also, 
   A. Kageyama, S. Kaneko, N. Shimoyama and M. Tanimoto, \Journal{\PLB}{538}{79}{2002};
   Z.Z. Xing, \Journal{\PLB}{539}{85}{2002}; \Journal{\PLB}{550}{178}{2002}.
\end{thebibliography}
\end{document}